\documentclass{aastex7}

\makeatletter
\newcommand*{\transpose}{%
  {\mathpalette\@transpose{}}%
}
\newcommand*{\@transpose}[2]{%
  \raisebox{\depth}{$\m@th#1\intercal$}%
}

\makeatother

\usepackage{amsmath,amssymb}
\usepackage{bm}

\begin{document}

\newcommand{\Teff}{T_{\mathrm{eff}}}
\newcommand{\logg}{\log{g}}
\newcommand{\vsini}{v \sin{i}}
\newcommand{\logRHK}{\log{R'_{\mathrm{HK}}}}
\newcommand{\SHK}{$S_{\mathrm{HK}}$}
\newcommand{\ms}{m$\,$s$^{-1}$}
\newcommand{\cms}{cm$\,$s$^{-1}$}
\newcommand{\kms}{km$\,$s$^{-1}$}
\newcommand{\Ha}{H$\alpha$}
\newcommand{\Kepler}{\emph{Kepler}}
\renewcommand*{\sectionautorefname}{Section}
\renewcommand*{\subsectionautorefname}{Section}
\renewcommand*{\subsubsectionautorefname}{Section}

\def\app#1#2{%
  \mathrel{%
    \setbox0=\hbox{$#1\sim$}%
    \setbox2=\hbox{%
      \rlap{\hbox{$#1\propto$}}%
      \lower1.1\ht0\box0%
    }%
    \raise0.25\ht2\box2%
  }%
}
\def\approxprop{\mathpalette\app\relax}

\newcommand{\sgnpz}{\mathrm{s}}

\definecolor{m1}{RGB}{0, 162, 255}
\definecolor{m2}{RGB}{0, 118, 186}
\definecolor{m3}{RGB}{0, 171, 142}
\definecolor{m4}{RGB}{29, 177, 0}
\definecolor{m5}{RGB}{1, 113, 0}

\definecolor{jkl}{rgb}{0.4, 0.0, 0.6}
\newcommand{\jkl}[1]{\textcolor{jkl}{\bf [JKL: #1]}}
\newcommand{\todo}[1]{\textcolor{jkl}{\bf [to do: #1]}}

\newcommand{\newtext}[1]{{\bf #1}}
\title{Exposure-averaged Gaussian Processes for Combining Overlapping Datasets}

\copyright\ 2026. All rights reserved. \\
\newcommand{\cca}{Center for Computational Astrophysics, Flatiron Institute, 162 Fifth Avenue, New York, NY 10010, USA}
\newcommand{\jpl}{Jet Propulsion Laboratory, California Institute of Technology, 4800 Oak Grove Drive, Pasadena, CA 91125, USA}
\newcommand{\columbia}{Department of Astronomy, Columbia University, 538 West 120th Street, Pupin Hall, New York, NY 10027, USA}
\newcommand{\uchicago}{Department of Astronomy \& Astrophysics, University of Chicago, 5640 S Ellis Avenue, Chicago, IL 60637, USA}

\author[orcid=0000-0002-4927-9925]{Jacob K. Luhn}
\affiliation{\jpl}
\email[show]{jacob.luhn@jpl.nasa.gov}

\author[0000-0003-3856-3143]{Ryan A. Rubenzahl}
\affiliation{\cca}
\email{rrubenzahl@flatironinstitute.org}

\author[0000-0003-1312-9391]{Samuel Halverson}
\affiliation{\jpl}
\email{}

\author[0000-0002-3852-3590]{Lily L. Zhao}
\thanks{NASA Sagan Fellow} 
\affiliation{\uchicago}
\email{lilylingzhao@uchicago.edu}

\begin{abstract}
Physically motivated Gaussian process (GP) kernels for stellar variability, like the commonly used damped, driven simple harmonic oscillators that model stellar granulation and p-mode oscillations, quantify the instantaneous covariance between any two points. For kernels whose timescales are significantly longer than the typical exposure times, such GP kernels are sufficient. For time series where the exposure time is comparable to the kernel timescale, the observed signal represents an exposure-averaged version of the true underlying signal. This distinction is important in the context of recent data streams from Extreme Precision Radial Velocity (EPRV) spectrographs like fast readout stellar data of asteroseismology targets and solar data to monitor the Sun's variability during daytime observations. Current solar EPRV facilities have significantly different exposure times per-site, owing to the different design choices made. Consequently, each instrument traces different binned versions of the same ``latent'' signal. Here we present a GP framework that accounts for exposure times by computing integrated forms of the instantaneous kernels typically used. These functions allow one to predict the true latent oscillation signals \emph{and} the exposure-binned version expected by each instrument. We extend the framework to work for instruments with significant time overlap (i.e., similar longitude) by including relative instrumental drift components that can be predicted and separated from the stellar variability components. We use Sun-as-a-star EPRV datasets as our primary example, but present these approaches in a generalized way for application to any dataset where exposure times are a relevant factor or combining instruments with significant overlap.

\end{abstract}



\section{Introduction}
Gaussian processes (GPs) have become a useful tool for characterizing and modeling stellar variability in astronomical time series (see \citealt{Aigrain2023} for a thorough review). A GP framework allows one to model stochastic signals in a flexible, nonparametric way via a fully Bayesian approach that properly propagates parameter uncertainties. The conditioned posterior specifies the prediction for the mean and Gaussian uncertainties for the latent process (i.e. the ``predictive mean'' and ``predictive variance'') given an observed dataset, providing a powerful framework for predicting future (or past) observations and interpolating between observations via informed models.

Modeling the observed manifestations of stellar variability as a GP largely stems from the desire to treat intrinsic stellar ``noise'' as a stochastic signal that complicates fitting an underlying model of interest, e.g., rotational modulation affecting exoplanet transit models (transit depth/ planet radius) \citep[e.g.][]{Carter2009,Gibson2012,Barclay2015}, or radial velocity (RV) curves \citep[e.g.][]{Aigrain2012,Haywood2014,Rajpaul2015}. In the era of extreme precision radial velocity (EPRV) surveys, using GPs to model stellar variability has become increasingly popular \citep{Gilbertson2021,Jones2022,Hara2025}. This is because stellar variability represents the largest remaining hurdle to routinely detecting and characterizing low-mass planets in the habitable zones of their host stars \citep{Crass2021} now that existing EPRV spectrographs are approaching or have achieved the required precision and stability \citep{Schwab2016,Gibson2016,Blackman2020,Pepe2021} for such detections in the absence of stellar variability \citep{Luhn2023}.
As a result, characterizing and mitigating stellar variability is an active area of research \citep{Crass2021,Zhao2022}. One challenging aspect of stellar variability is the multitude of astrophysical sources, each with their own amplitude and timescale \citep[e.g.,][]{Luhn2020a}. While sources of variability that operate on timescales longer than a day (e.g., rotational modulation from spots/faculae, years-long activity cycles) can be reasonably sampled with existing RV surveys, sources of variability that operate on daily timescales or less (e.g., 12-24 hr supergranulation, hours-long granulation, and minutes-long p-mode oscillations) require significantly higher-cadence observations, which becomes unfeasible for large-scale surveys that target an appreciable number of stars \citep{Newman2023,Luhn2023}.

To combat the challenges of stellar variability, many existing EPRV spectrographs are now equipped with a solar feed that obtains daytime Sun-as-a-star observations \citep{Dumusque2015,Lin2022,Rubenzahl2023,Llama2024}. The solar feeds operate ``for free'' during the day and, since we can independently precisely and accurately remove the RV signals due to solar system planets, provide high-cadence time series of purely solar variability and instrumental drift. Consequently, these data streams do not suffer from the same limitations in terms of cadence as nighttime data, and as a result have been extremely valuable for studying solar signals \citep[see, e.g.,][]{Milbourne2019,Haywood2022,Ford2024,O'Sullivan2025}.
Solar feeds now exist on the High-Accuracy Radial-velocity Planet Searcher (HARPS), the High-Accuracy Radial-velocity Planet Searcher in the North \citep[HARPS-N;][]{Phillips2016}, the EXtreme PREcision SPectrograph \citep[EXPRES;][]{Llama2024}, the NEID spectrograph \citep{Lin2022}, and the Keck Planet Finder \citep[KPF;][]{Rubenzahl2023}.  The earliest solar telescope on-sky, feeding into HARPS-N, was commissioned in 2015.  Typical exposure times for the different solar feeds can vary from five seconds to over five minutes.

The solar datasets are even more useful in combination. The longitudinal distribution of existing solar telescopes results in a 15-hour continuous segment of solar data from UT0900 (HARPS-N) to UT2400 (EXPRES) in the summer, with multiple instances of overlapping coverage between instruments\footnote{In the winter, the continuous segment is reduced to 11 hours from UT1100 (HARPS-N) to UT2200 (KPF). Note that KPF does not currently operate in the afternoon; extending KPF solar observations into the afternoons would result in 19-hour segments in the summer and 15-hour segments in the winter.}. In some cases, instruments at nearly the same longitude yield completely overlapping datasets, though of course the continuous coverage and overlap will vary day to day due to weather conditions at each location. The overlap between successive and co-located instruments adds an additional use case beyond studying stellar variability: the ability to perform cross-instrument comparisons. \citet{Zhao2023} provided the first attempt to combine solar data from multiple instruments to perform such cross-comparisons. Another recent cross-instrument comparison in \citet{Rubenzahl2023} examined the newly commissioned KPF/SoCal observations against NEID. 

However, these combined analyses require some care, as not all spectrographs observe with the same cadence and exposure times. HARPS-N and EXPRES observations have exposure times on the order of minutes (roughly 5 and 3 minutes respectively) while HARPS, NEID, and KPF have sub-minute exposure times (30~s, 55~s, and 12~s respectively).  To address the different exposure times and time stamps, \citet{Zhao2023} binned all observations onto common time stamps with 16.2-minute-wide bins spaced by 5.4 minutes. As a result, the time series of the combined data set was insensitive to the shortest-period sources of variability, namely p-mode oscillations and the high-frequency tail of granulation. Given the higher cadence and shorter exposures of KPF and NEID, \citet{Rubenzahl2023} used a cubic Hermite spline to interpolate the faster-cadence KPF data at the NEID timestamps to compute relative residuals between instruments. While sufficient for comparing instrument performance between KPF and NEID, such interpolation schemes are not informed by physically motivated processes, nor can they be extended to compare two data sets with significantly different exposure times, where each instrument will trace different exposure-averaged manifestations of the true underlying signal.

Here we aim to build on these previous two analyses with a more robust framework that leverages the capabilities of GP modeling. We will start with physically motivated kernels for stellar variability that have been used extensively in the literature \citep{ForemanMackey2017,Pereira2019,Guo2022,Luhn2023,Gupta2024a}. Importantly, these GP kernels represent the \emph{instantaneous} covariance functions, and thus do not contain exposure-time information. Naively combining datasets from multiple instruments with different exposure times using such kernels will introduce spurious signals, since the GP model cannot reconcile two closely spaced observations that have significant, real RV differences from their different exposure times. Instead, we will build a GP framework from the integrated GP kernels that properly accounts for the effects of finite exposure times to predict the common latent process uniquely traced by each instrument. 

The manuscript is laid out as follows. In \autoref{sec:gp_background}, we provide an overview of GPs and their limitations when considering exposure-time effects. In \autoref{sec:integration_framework}, we describe the ingredients necessary to build a GP framework that accounts for exposure-time effects, and provide the analytic functions for the integrated forms of the typical damped and driven Simple Harmonic Oscillator (SHO) kernels used to model astrophysical stellar granulation and p-mode oscillations. \autoref{sec:gp_decomposition} reviews the methodology for decomposing a multicomponent GP model into its individual components. Building toward a full GP model to combine solar EPRV datasets from multiple instruments, \autoref{sec:instrument_drift} introduces the framework for accounting for instrumental drifts between each spectrograph.  We put these pieces together into a full model with synthetic data as a proof of concept in \autoref{sec:synthetic_test}, and in \autoref{sec:real_test} we apply it to two cases of real solar data. We finish with some possible applications of exposure-time-informed GP kernels to other domains in \autoref{sec:broader_applications}, discuss current limitations and next steps in \autoref{sec:discussion}, and summarize our conclusions in \autoref{sec:conclusions}.

\section{Gaussian Process Notation}\label{sec:gp_background}
A Gaussian process is a stochastic process where the joint probability distribution of a set of 
$\bm{y} = \left\{y_i\right\}_{i=1...N}$ evaluated at input locations $\bm{x} = \left\{x_i\right\}_{i=1...N}$ is a Gaussian distribution
\begin{equation}
p(\bm{y}) = \mathcal{N}(\bm{m},\mathbf{K}),
\end{equation} where $\bm{m}$ is the mean vector given by the mean function ${m}_i = m(x_i,\bm{\theta})$ and $\mathbf{K}$ is the covariance matrix defined by the kernel function $K_{i,j} = k(x_{i},x_{j},\bm{\phi})$. 
In general the inputs $x_i$ can represent spatial or temporal coordinates; because we analyze time series data in this work, we use $t$ in place of $x$ throughout. While the mean function $m$ can be any function, the kernel function $k$ and its hyperparameters $\bm \phi$ are what define the GP; considerable care must be taken when choosing a kernel function, as it must produce a positive definite covariance matrix. As such, one typically chooses from a family of established kernel functions (squared exponential, periodic, Matérn, etc.). Conveniently, any affine transformation of a positive-definite kernel results in another positive-definite kernel, so custom kernels can be created from sums and products of these ``base'' kernels.
The mean function, $\bm{m}$ can often be a constant or even zero (e.g., ``zero mean'' when modeling data solely with a GP). However, GPs are also typically used to model a ``nuisance signal'' (or a sum of multiple such signals) that contaminates the actual signal of interest \citep{Rasmussen2006}, such as the case of modeling stellar variability as a GP when fitting a Keplerian model to RV data. In that case, the Keplerian model is the mean function.

In addition to using GP regression to estimate the hyperparameters of the mean and kernel functions ($\bm{\hat{\theta}}$, $\bm{\hat{\phi}}$, respectively), GPs allow us to predict behavior at any set of test points using conditional probabilities. Using the example of time series data, given a set of noisy observations $\bm{y}$ at times $\bm{t}$ with measurement uncertainties $\bm{\sigma}$, the behavior at a new set of times, $t_\ast$, is a Gaussian 
\begin{equation}
    \bm{y}_{{\ast}} \mid \bm{y} = \mathcal{N} (\bm{\mu}_{\rm{\ast}},\mathbf{C}_{\rm{\ast}})
\end{equation}
with mean 
\begin{equation}
    \bm{\mu}_{\rm{\ast}} = \bm{m}_{\rm{\ast}} + \mathbf{K}_{\rm{\ast}}^{\transpose} 
    \left(\mathbf{K}+\mathbf{R}\right)^{-1} (\bm{y} - \bm{m})
    \label{eqn:predictive_mean}
\end{equation}
and covariance
\begin{equation}
    \mathbf{C}_{\ast} = \mathbf{K}_{\ast \ast} -  \mathbf{K}_{\rm{\ast}}^{\transpose} 
    \left(\mathbf{K}+\mathbf{R}\right)^{-1} 
    \mathbf{K}_{\rm{\ast}},
    \label{eqn:conditioned_covar}
\end{equation}
 where $\mathbf{R}$ is the noise matrix containing the measurement variances and is most often diagonal, (e.g., $\mathbf{R} \equiv \bm{\sigma} \mathbf{I}$). Here we have used $\ast$ notation such that $\bm{m}_{\ast} \equiv m(\bm{t}_{\ast},\bm{\theta})$, that is, the mean function evaluated at test points $\bm{t_{\ast}}$, and $\mathbf{K}_{\ast}$ is the covariance between input observations and the test points
 \begin{equation}
     \left[K_{\ast}\right]_{i,j} \equiv k(t_{i},t_{\ast,j}; \bm{\phi}).
 \end{equation}
 Likewise $\mathbf{K}_{\ast\ast}$ is the covariance between all pairs of test points
 \begin{equation}
     \left[K_{\ast\ast}\right]_{i,j} \equiv k(t_{\ast,i},t_{\ast,j}; \bm{\phi}).
 \end{equation}

\section{Kernel Integration Framework}\label{sec:integration_framework}
\subsection{Requisite Kernel Integrations}
Here we closely follow the steps outlined by \citet{Smith2018} to describe the ingredients necessary for a Gaussian process framework to predict a latent function from binned data, with a key difference that in our case, the binning represents an exposure-averaged signal rather than a strict sum. We begin with the knowledge that our observed data come from an integration over a latent function $f(t)$, which we wish to predict. We assume a function $F(t_1,t_2)$ describes the integral of $f(t)$ between time $t_1$ and $t_2$, which for example represent the start and stop times of an exposure. Then our observations $y(t_1,t_2)$ are noisy samples of $F(t_1,t_2)$. We can therefore construct a Gaussian process from which $f(t)$ and $F(t_1,t_2)$ are jointly drawn. The Gaussian process posterior can be constructed for expressions for the covariance between $f(t)$ and $f(t')$ (covariance of the latent function), $F(t_1,t_2)$ and $F(t_1',t_2')$ (the covariance between integrated observations), and the cross-covariance between the latent function and the integrated observation $F(t_1,t_2)$ and $f(t')$.

For now, we ignore the exact form of the kernel function and describe the requisite integrals needed. We assume we have a GP kernel to describe the `latent' function $k_{ff}(t,t')$. Analogous to \citet{Smith2018}, our observations, which come from $F(t_1,t_2) = \frac{1}{t_2 - t_1}\int_{t_1}^{t_2}f(t)dt$ is also a GP with covariance described by the integrated latent kernel
\begin{equation}\label{eqn:double_int_general}
    k_{FF}((t_1,t_2),(t_1',t_2')) = \frac{1}{\left({t_2}-{t_1}\right)\left({t_2'}-{t_1'}\right)}\int_{t_1}^{t_2}\int_{t_1'}^{t_2'}k_{ff}(t,t')dt' dt.
\end{equation}
where the normalization factor out front reflects the exposure-averaged covariance\footnote{This definition assumes a uniform photon arrival rate. Exploring time-variable photon arrival rates would require numerical integration and are beyond the scope of this paper.}. The cross covariance between an observation and the latent function is similarly
\begin{equation}\label{eqn:single_int_general}
    k_{Ff}((t_1,t_2),(t')) = \frac{1}{\left({t_2}-{t_1}\right)} \int_{t_1}^{t_2}k_{ff}(t,t')dt.
\end{equation}
With the necessary covariance functions defined, we can evaluate the covariance matrix in the case of a latent function conditioned on a set of exposure-integrated observations. In the next sections, we will outline an algorithm for evaluating $k_{FF}$ and $k_{Ff}$ for overlapping exposures and provide analytic expressions for these functions in the case of kernels that describe stellar oscillations and granulation.

\subsection{Sub-exposure decomposition}
Before we derive the integrated kernel functions for specific latent kernels, it is important to consider how we plan to evaluate \autoref{eqn:double_int_general} \& \autoref{eqn:single_int_general} for a given set of observations. The most common GP kernels are stationary; they only depend on the absolute time difference $|t-t'|$. We must therefore handle the integrals with care to account for the sign flip between $t<t'$ and $t>t'$. These special cases occur any time we wish to evaluate the integral for a time that is within one of the observations, for example evaluating $k_{FF}$ in the case of two observations with partial or complete overlap, or when evaluating $k_{Ff}$ to predict the latent function during one of the observations. We note that the case of overlapping observations will always occur when evaluating the diagonal of the covariance matrix (the covariance between each observation and itself) even in a time series of observations from a single data stream of sequential observations. 

\begin{figure}
    \centering
    \includegraphics[width=\linewidth]{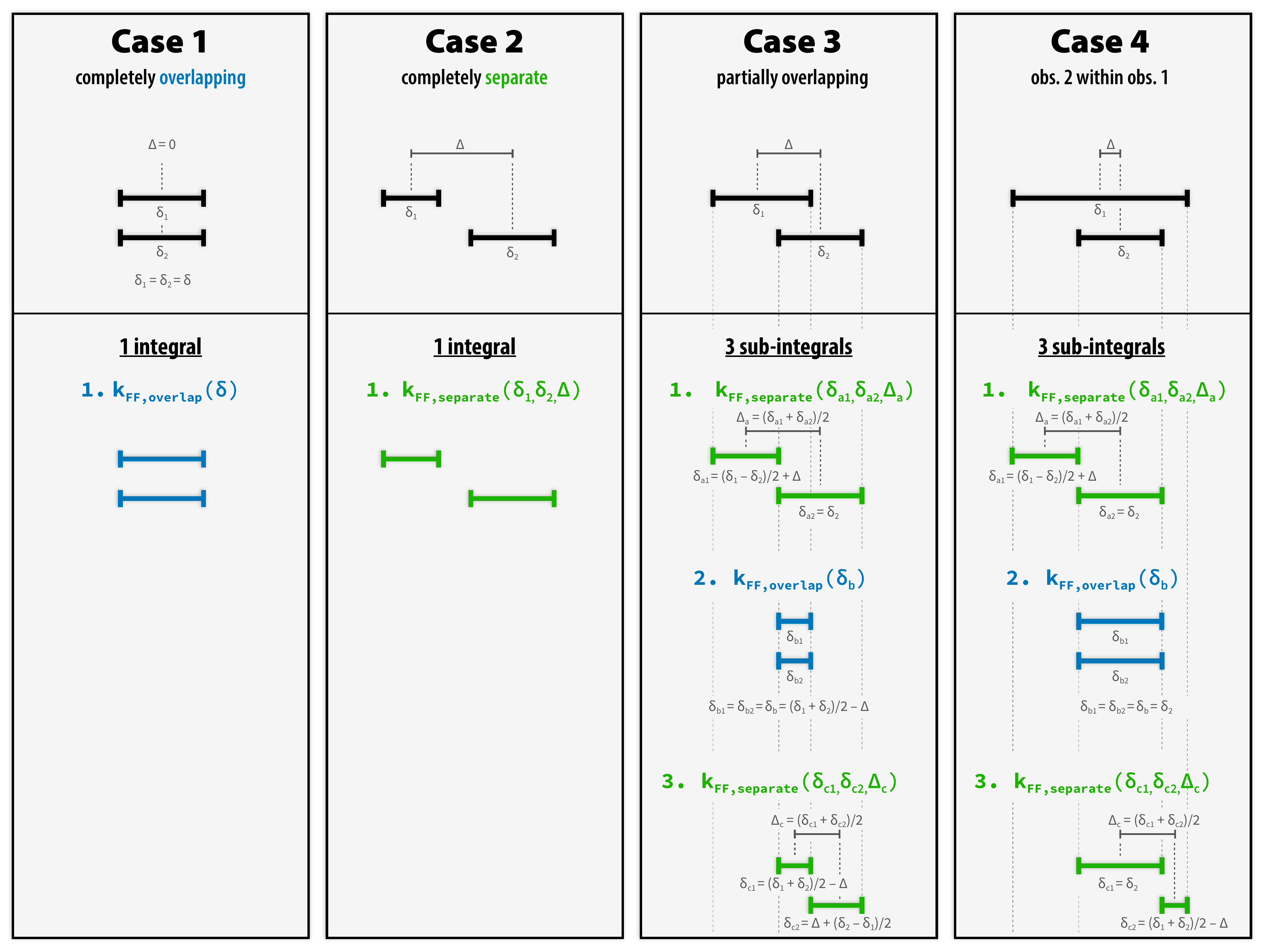}
    \caption{When computing the double integral $k_{FF}$, any two exposures can be decomposed into a sum of perfectly overlapping and separate sub-exposures; each sub-integral is evaluated with newly defined exposure times ($\delta$s) and time separations ($\Delta$s). Note that to compute the sum of the three sub-integrals, one must first multiply each sub-integral by the product of the two sub-exposure times before summing, and finally dividing by the original exposure time product to obtain the final result. For example, in Case 3, the total integral $k_{FF}(\delta_1,\delta_2,\Delta) = 1/({\delta_1\delta_2}) \left(k_{FF,sep}(\delta_{a1},\delta_{a2},\Delta_a)*\delta_{a1}\delta_{a2} + k_{FF,over}(\delta_{b})*\delta_{b}^2 + k_{FF,sep}(\delta_{c1},\delta_{c2},\Delta_c)*{\delta_{c1}\delta_{c2}}\right)$.}
    \label{fig:overlap_decompose}
\end{figure}

The double-integrated covariance $k_{FF}$ of a stationary kernel is simply a function of the two exposure durations, $\delta_i$ and $\delta_j$ (i.e., the integration times of the two exposures), and the time separation between observations $\Delta$, which in our reference frame is defined as the difference between the two exposure midpoints. We define $k_{FF,\rm{overlap}}(\delta_i=\delta_j\equiv \delta,\Delta=0)$ and $k_{FF,\rm{separate}}\left(\delta_i,\delta_j,\Delta > (\delta_i+\delta_j)/2\right)$ as the functions describing completely overlapping and completely separate observations, respectively. Still assuming stationary kernels, $k_{FF}$ is invariant to $t \leftrightarrow t'$, and so for simplicity, we redefine $i$ and $j$ such that $\delta_i > \delta_j$ (i.e., observation $i$ is the longer of the two exposures), and we define a coordinate system such that observation $i$ begins at time $t=0$ (and therefore ends at $t=\delta_i$). Any two observations $i,j$ will therefore fall into 1 of 4 cases: 1) completely overlapping, 2) completely separate, 3) partially overlapping, 4) observation $j$ is within observation $i$. The full integral $k_{FF}$ for each of these cases can be decomposed into a sum of $k_{FF,\rm{overlap}}$ and $k_{FF,\rm{separate}}$, as illustrated in \autoref{fig:overlap_decompose}.

The single integral $k_{Ff}$ follows a similar approach, however in this case we only need to derive one integrated function, and we have chosen to assume $t>t'$. We again note that it is $t\leftrightarrow t'$ invariant, and thus, the single integrated function only depends on the exposure time $\delta$, and a time separation (the difference between the center of the observation and the time at which the latent kernel is being evaluated), $\Delta$. If evaluating $k_{Ff}$ for a time $t'$ within an exposure, the integral is simply split into two sub-integrals at $t'$, redefining $\delta$ and $\Delta$ for the left and right portions in turn, and summing each portion.

\subsection{Integrating the SHO Kernel}
Here we provide the analytic expressions for the integrated kernels where the latent kernel is a stochastically driven, damped simple harmonic oscillator (SHO). This choice of kernel is ideal for modeling convectively driven stellar processes like granulation and p-mode oscillations \citep{ForemanMackey2017,Pereira2019,Luhn2023}. In particular, when the oscillator strength is sufficiently large, ($Q>1/2$), the kernel takes the form:
\begin{equation}
    k_{ff}^{\rm{SHO}}(|t-t'|) = S_0 \omega_0 Q \, e^{-\frac{\omega_0 |t-t'|}{2 Q}} \left(\cos{\left(\eta \omega_0 |t-t'|\right)} + \frac{1}{2 \eta Q}\sin{\left(\eta \omega_0 |t-t'|\right)}\right)
\end{equation}
where $\omega_{0}$ is the characteristic frequency of the undamped oscillator, $S_{0}$ is proportional to the power of the spectrum at $\omega = \omega_{0}$, $Q$ is the quality factor, and $\eta = |1-(4Q^2)^{-1}|^{1/2}$.

For the cross-covariance between an observation with exposure time $\delta$ and a test point $t'$ separated from the center of the exposure by $\Delta$ ($t'\equiv \Delta + \delta/2$), it is easiest to define the coordinate system such that $t'=0$ and $t-t' > 0$ (i.e., the test point is at the origin and the exposure being integrated over spans positive times. The single integral of the latent kernel is then
\begin{equation}
    k_{Ff}^{\rm{SHO}} (\delta,\Delta) = \frac{1}{\delta}\int_{\Delta -\frac{\delta}{2}}^{\Delta + \frac{\delta}{2}} k_{ff}^{\rm{SHO}}\left(\left|t-t'\right|\rightarrow t\right)\,dt = \left(\frac{S_0 Q}{\delta \eta }\right)\left(\frac{1}{1+a^2}\right) \,  \, I_0^{\rm{SHO}}(u_1,u_2), 
    \label{eqn:sho_single_int}
\end{equation}
where $a = 1/(2Q\eta)$ and we have defined the helper equation
\begin{equation}
    I_0^{\rm{SHO}}(u_a,u_b) = e^{-au} \left[\left(1-a^2\right)\sin{u} - 2a\cos{u}\right]\Big|_{u=u_a}^{u=u_b}
\end{equation}
with bounds
\begin{equation}
    u_1 = \eta\omega_0\left(\Delta - \frac{\delta}{2}\right) \quad \text{and} \quad u_2 = \eta\omega_0\left(\Delta +\frac{\delta}{2}\right)
    \label{eqn:sho_single_int_bounds}
\end{equation}
We note that \autoref{eqn:sho_single_int} and associated bounds (\autoref{eqn:sho_single_int_bounds}) can also be obtained by evaluating the equivalent ``negative" integral $\frac{1}{\delta}\int_{-\Delta -\frac{\delta}{2}}^{-\Delta + \frac{\delta}{2}} k_{ff}^{\rm{SHO}}\left(\left|t-t'\right|\rightarrow -t\right)\,dt$. However, both ``positive" and ``negative" expressions only hold for cases where $\Delta > \delta/2$ (i.e., the test point is not within the exposure), as a result of our assumption $t-t' > 0$ ($t-t' < 0$ in the ``negative" case) required to evaluate the integral analytically. To handle the case of a test point that lies within an observation, we break the integral into two subintegrals on either side of the test point with new durations $\delta_a$, $\delta_b$ and separations $\Delta_a$ and $\Delta_b$. The full integral is renormalized by $\delta^{-1} \left(k_{Ff}^{\rm{SHO}}(\delta_a,\Delta_a)*\delta_a  + k_{Ff}^{\rm{SHO}}(\delta_b,\Delta_b)*\delta_b\right)$.

The double integral of the latent kernel was derived in \citet{Luhn2023}, which we repeat here for completeness. In the case of perfectly overlapping observations, 
\begin{equation}
    k_{FF,\rm{overlap}}^{\rm{SHO}} (\delta) = \frac{1}{\delta^2}\int_{0}^{\delta}dt' \int_{0}^{\delta} k_{ff}^{\rm{SHO}}(t,t') \, dt =  \frac{2 S_0 Q}{\delta^2 \eta \left(1+a^2\right)} \Big(I_{1}^{\rm{SHO}}\left(u_1,u_2\right) - I_{2}^{\rm{SHO}}\left(u_1,u_2\right) 
+ 2 a \delta \Big)
\end{equation}
using helper functions
\begin{equation}
I_{1}^{\rm{SHO}}\left(u_a,u_b\right) = \frac{e^{-au}\left(1-a^2\right)}{\eta \omega_{0} \left(1+a^2\right)} \left(\cos{u} + a\sin{u} \right) \Big|_{u=u_a}^{u=u_b}
\label{eqn:I1}
\end{equation}
and
\begin{equation}
I_{2}^{\rm{SHO}}\left(u_a,u_b\right) = -\frac{2a e^{-au}}{\eta \omega_{0} \left(1+a^2\right)} \left(\sin{u} - a\cos{u} \right) \Big|_{u=u_a}^{u=u_b}
\label{eqn:I2}
\end{equation}
with bounds
\begin{equation}
    u_1 = \eta\omega_0\delta \quad \text{and} \quad u_2 = 0
\end{equation}

For two completely separate observations, 
\begin{equation}
\begin{split}
    k_{FF,\rm{separate}}^{\rm{SHO}} (\delta_1,\delta_2,\Delta) &= \frac{1}{\delta_1 \delta_2} \int_0^{\delta_1}dt' \int_{\frac{\delta_1-\delta_2}{2}+\Delta}^{\frac{\delta_1+\delta2}{2}+\Delta} k_{ff}^{\rm{SHO}}(t,t')\, dt \\
    &= \frac{S_0 Q }{\delta_1 \delta_2 \eta \left(1+a^2\right)} \Big(I_{1}^{\rm{SHO}}\left(u_1,u_2\right) - I_{2}^{\rm{SHO}}\left(u_1,u_2\right) 
- I_{1}^{\rm{SHO}}\left(u_3,u_4\right) + I_{2}^{\rm{SHO}}\left(u_3,u_4\right)\Big) 
\label{eqn:osc_double_integral}
\end{split}
\end{equation}
using the same $I_1$ and $I_2$ as above for the overlapping integral, but with bounds
\begin{equation}
\begin{split}
u_1 = \eta \omega_{0} \left(\frac{\delta_1 + \delta_2}{2} + \Delta\right), \quad \quad u_2 = \eta \omega_{0} \left(\Delta - \frac{\delta_1 - \delta_2}{2}\right), \quad \quad \\
u_3 = \eta \omega_{0} \left(\frac{\delta_1 - \delta_2}{2} + \Delta\right), \quad \quad u_4 = \eta \omega_{0} \left(\Delta - \frac{\delta_1+\delta_2}{2}\right). \quad \quad 
\end{split}
\end{equation}

\subsubsection{Stellar p-mode oscillations}
The kernel integrals given above can be used for modeling p-mode oscillations in a few different ways. \citet{ForemanMackey2017} used a sum of SHO kernels with very large $Q$ to model individual modes of oscillations, imitating the comb-like structure of Lorentzian modes observed in asteroseismic analyses of long-baseline time series. \citet{Pereira2019} and \citet{Luhn2023} instead used a single SHO kernel to model these modes as a Gaussian-like power envelope centered on the frequency of maximum power, $\nu_{max}$, as asteroseismic scaling relations exist for both the power and frequency of maximum oscillations. We will adopt the latter approach for its simplicity, but also because the time series we will model will not be sufficiently long enough to resolve individual modes of oscillation. Further, by properly computing the integrated kernels, the kernels are no longer strictly stationary, since they depend not just on $\Delta \equiv \left| t-t'\right|$, but also on the individual exposure times $\delta_1$ and $\delta_2$. Specifically in the case of overlapping measurements, the logic to decompose overlaps does not have an exploitable structure in the covariance matrix, and so to our knowledge we cannot make use of the scaling improvements outlined in \citet{ForemanMackey2017} that enable efficient analyses of long-baseline time series with $N_{obs}\gtrsim 1000$.

\subsubsection{Stellar granulation}
The PSD for a SHO kernel with $Q=1/\sqrt{2}$ closely follows a Harvey-like \citep{Harvey1985} function, and so it is typically used to model granulation components \citep{ForemanMackey2017,Pereira2019,Luhn2023}. Since $1/\sqrt{2} > 1/2$, the equations above are also valid for granulation with $Q$ fixed at $1/\sqrt{2}$. Granulation is often modeled as a sum of two such kernels \citep{Guo2022,Luhn2023}, or more to also include lower-frequency supergranulation structure \citep{OSullivan2024}. We will adopt a 2-component model in this work, since the time series we will model will not be long enough to probe supergranulation timescales (12+ hours).

\section{Decomposing Multi-component GP Models}\label{sec:gp_decomposition}
The model we will employ for stellar variability will be a sum of several independent GPs, for example, 
\begin{equation}
    f_{\star}(t) = f_{osc}(t) + f_{gran}(t) + f_{supergran}(t) + f_{rot}(t) \ +\   ...   = \sum_{l=1}^{N_l}f_l(t),
\label{eqn:stellar_gp}
\end{equation}
where the subscript $\star$ (or $\odot$ in the case of the Sun) refers specifically to the ``latent" stellar components, and the far right-hand side is a generalized sum over $N_l$ latent stellar processes. We therefore wish to describe the procedure we will use throughout to evaluate the predictive mean of each component in such a multi-component GP model. Following a similar notation to \citet{Duvenaud2014}, the standard GP result for how to condition data (i.e., compute the predictive mean) based on a set of observations shows that if
\begin{equation}
    \begin{bmatrix} \bm{y}_A \\ \bm{y}_B \end{bmatrix} \sim N\left(\begin{bmatrix} {\bm{m}_A} \\ {\bm{m}_B} \end{bmatrix}, \begin{bmatrix} \mathbf{K}_{AA} & \mathbf{K}_{AB} \\ \mathbf{K}_{BA} & \mathbf{K}_{BB}
    \end{bmatrix}
    \right).
    \label{eqn:general_gp_distribution}
\end{equation}
then
\begin{equation}
    \bm{y}_A\mid\bm{y}_B \sim N\left(\bm{m}_A + \mathbf{K}_{AB}\left(\mathbf{K}_{BB}+\mathbf{R}\right)^{-1} \left(\bm{y}_B-\bm{m}_B\right), \mathbf{K}_{AA} - \mathbf{K}_{AB} \left(\mathbf{K}_{BB}+\mathbf{R}\right)^{-1} \mathbf{K}_{BA}\right),
    \label{eqn:general_gp_prediction}
\end{equation}
where $\mathbf{R}$ is the noise matrix containing the measurement variances, $\bm{y}_B$ represents a set of observations observed at times $\{t_1,t_2,...t_N\}$, and $\bm{y}_A$ represents a set of test points over which to evaluate the predictive distribution, $\{t_{\ast,1},t_{\ast,2},...t_{\ast,M}$\}. We have included the means $\bm{m}_A$ and $\bm{m}_B$ for completeness, though these are often simply 0. The relevant covariance matrices are
\begin{equation}
    \begin{aligned}
        \mathbf{K}_{AA} &= k(\bm{t_{\ast}},\bm{t_{\ast}}) \quad  &
        \mathbf{K}_{AB} &= k(\bm{t_{\ast}},\bm{t}) \\
        \mathbf{K}_{BA} &= k(\bm{t},\bm{t_{\ast}}) \quad &
        \mathbf{K}_{BB} &= k(\bm{t},\bm{t}).
    \end{aligned} 
\end{equation}

For a simple case of two independent GPs, $f_1 \sim GP(m_1,K_1)$ and $f_2 \sim GP(m_2,K_2)$, then
\begin{equation}
    \begin{bmatrix} \bm{f}_1(x) \\ \bm{f}_1(x_{\ast}) \\ \bm{f}_2(x) \\ \bm{f}_2(x_{\ast}) \\ \bm{f}_1(x) + \bm{f}_2(x)  \\ \bm{f}_1(x_{\ast}) + \bm{f}_2(x_{\ast})\end{bmatrix} \sim N\left(\begin{bmatrix} {\bm{m}_1} \\ {\bm{m}_{1,\ast}} \\ {\bm{m}_2} \\ {\bm{m}_{2,\ast}} \\ {\bm{m}_1} + {\bm{m}_2}\\ {\bm{m}_{1,\ast} + \bm{m}_{2,\ast}}\end{bmatrix}, \begin{bmatrix} \mathbf{K}_{1} & \mathbf{K}_{1,\ast} & 0 & 0 & \mathbf{K}_{1} & \mathbf{K}_{1,\ast} \\ 
    \mathbf{K}_{1,\ast}^{\transpose} & \mathbf{K}_{1,\ast\ast} & 0 & 0 & \mathbf{K}_{1,\ast} & \mathbf{K}_{1,\ast\ast} \\
    0 & 0 & \mathbf{K}_{2} & \mathbf{K}_{2,\ast} & \mathbf{K}_{2} & \mathbf{K}_{2,\ast} \\
    0 & 0 & \mathbf{K}_{2,\ast}^{\transpose} & \mathbf{K}_{2,\ast\ast} & \mathbf{K}_{2,\ast} & \mathbf{K}_{2,\ast\ast} \\
    \mathbf{K}_{1} & \mathbf{K}_{1,\ast}^{\transpose} & \mathbf{K}_{2} & \mathbf{K}_{2,\ast}^{\transpose} & \mathbf{K}_{1} + \mathbf{K}_{2} & \mathbf{K}_{1,\ast} + \mathbf{K}_{2,\ast}\\
    \mathbf{K}_{1,\ast}^{\transpose} & \mathbf{K}_{1,\ast\ast} & \mathbf{K}_{2,\ast}^{\transpose} & \mathbf{K}_{2,\ast\ast} & \mathbf{K}_{1,\ast} + \mathbf{K}_{2,\ast} & \mathbf{K}_{1,\ast\ast} + \mathbf{K}_{2,\ast\ast}
    \end{bmatrix}
    \right),
    \label{eqn:independent_gp_sum}
\end{equation}
where $\mathbf{K}_{p,\ast}$ is the covariance matrix for a single process $p$ evaluated between the input observations and the test points, and $\mathbf{K}_{p,\ast \ast}$ is the same for the covariance between all pairs of test points. Equations \ref{eqn:general_gp_distribution} and \ref{eqn:general_gp_prediction} then give the conditional probability and the predictive mean of one component process given the observed data
\begin{equation}
    \begin{bmatrix} \bm{y}_{1,\ast} \\ \bm{y}_1 + \bm{y}_2 \end{bmatrix} \sim N\left(\begin{bmatrix} {\bm{m}_{1,\ast}} \\ \bm{m}_1 + \bm{m}_2  \end{bmatrix}, \begin{bmatrix} \mathbf{K}_{1,\ast\ast} & \mathbf{K}_{1,\ast} \\ \mathbf{K}_{1,\ast}^{\transpose} & \mathbf{K}_1 + \mathbf{K}_2
    \end{bmatrix}
    \label{eqn:two_component_contitional}
    \right).
\end{equation}
\begin{equation}
    \begin{aligned}
    \bm{y}_{1,\ast} \mid \bm{y_1} + \bm{y_2} &\sim N\left(\bm{\mu}_{1,\ast}, \mathbf{C}_{1,\ast}\right) \\
    \bm{\mu}_{1,\ast} &= 
    \bm{m}_{1,\ast} + \mathbf{K}_{1,\ast}^{\transpose} \left(\mathbf{K}_1 + \mathbf{K}_2+  \mathbf{R}\right)^{-1} \left(\bm{y}_1 + \bm{y}_2  -  \bm{m}_1 - \bm{m}_2 \right)   \\
    \mathbf{C}_{1,\ast} &= \mathbf{K}_{1,\ast \ast} - \mathbf{K}_{1,\ast}^{\transpose} \left(\mathbf{K}_1 + \mathbf{K}_2 + \mathbf{R}\right)^{-1} \mathbf{K}_{1,\ast},
    \end{aligned}
    \label{eqn:two_component_mean}
\end{equation}
where we have again used $\bm{\mu}$ and $\mathbf{C}$ to denote the mean and covariance of a GP that has been \emph{conditioned} on a given set of data.

In the more general case where a GP model is built from a sum of $N_{p}$ independent GPs, the covariance, and mean functions become
\begin{equation}
\begin{aligned}
    \mathbf{K} &= \mathbf{K}_1 + \mathbf{K}_2 + \ldots + \mathbf{K}_{N_p} = \sum_{p=1}^{N_p} \mathbf{K}_p, \\
    \bm{m} &= \bm{m}_1 + \bm{m}_2 + \ldots + \bm{m}_{N_p} = \sum_{p=1}^{N_p} \bm{m}_p
\end{aligned}
\end{equation}
the $p$th component process along the predicted points $\{t_{\ast,1},t_{\ast,2},...,t_{\ast,M}\}$ and observations of the total sum (the generalized version of \autoref{eqn:two_component_contitional}) are distributed as
\begin{equation}
    \begin{bmatrix} \bm{y}_{p,\ast} \\ \bm{y} \end{bmatrix} \sim N\left(\begin{bmatrix} {\bm{m}_{p,\ast}} \\ \bm{m} \end{bmatrix}, \begin{bmatrix} \mathbf{K}_{p,\ast\ast} & \mathbf{K}_{p,\ast} \\ \mathbf{K}_{p,\ast}^{\transpose} & \mathbf{K}
    \end{bmatrix}
    \right), 
\end{equation}
and the mean of the $p$th process (the generalized version of \autoref{eqn:two_component_mean}) can be predicted via
\begin{equation}
    \begin{aligned}
    \bm{y}_{p,\ast} \mid \bm{y} &\sim N\left(\bm{\mu}_{p,\ast}, \mathbf{C}_{p,\ast}\right) \\
    \bm{\mu}_{p,\ast} &= 
    \bm{m}_{p,\ast} + \mathbf{K}_{p,\ast}^{\transpose} \left(\mathbf{K} + \mathbf{R}\right)^{-1} \left(\bm{y} -  \bm{m} \right)   \\
    \mathbf{C}_{p,\ast} &= \mathbf{K}_{p,\ast \ast} - \mathbf{K}_{p,\ast}^{\transpose} \left(\mathbf{K} + \mathbf{R}\right)^{-1} \mathbf{K}_{p,\ast},
    \end{aligned}
\end{equation}

When comparing the predictive mean of a component process $\mu_{\ast,p}$ to the predictive mean from the full model in \autoref{eqn:predictive_mean}, the only practical difference is that $\mathbf{K}^{\transpose}_{\ast,p}$ only contains the covariance due to a single component kernel; a similar statement for $\mathbf{K}_{\ast \ast,p}$, $\mathbf{K}^{\transpose}_{\ast,p}$, and $\mathbf{K}_{\ast,p}$ is true for the conditioned covariance.

Finally, we note that the associative property of addition makes the GP model flexible to arbitrary groupings of kernels. This is handy when individual astrophysical sources may be best represented with multiple components themselves, e.g., using multiple SHO kernels at the rotation period and its harmonics for a single ``rotation'' kernel, or as we will describe in the next section, grouping all stellar components together to separate ``instrument" from ``stellar" components.

\section{Multi-instrument GP Framework and Full Model}\label{sec:instrument_drift}
In the previous sections, we outlined how to account for finite exposure times to predict latent functions and reviewed how to compute the conditional predictive means for individual components of a multi-component GP model. Later, we will apply these radial velocity time series of the Sun from multiple instruments. Given that these datasets often have significant overlap, it will be useful to construct a GP model that treats each spectrograph's instrumental drifts as additional independent GPs, which we build as a multivariate GP where each instrument is its own parallel time series. Thus, for $N_s$ spectrographs\footnote{A more general term ``instruments" could be used throughout, however, we opt for ``spectrographs" as it reflects our motivating example and allows us to use a distinguishable iterator $s$.} if each spectrograph's GP is $f_s(t)$, the observed time series $q_s(t)$ are modeled for each instrument $s$ as 
\begin{equation}
\begin{aligned}
    q_1(t) &= f_1(t) + f_{\star}(t) , \\
    q_2(t) &= f_2(t) +  f_{\star}(t), \\
    \vdots \\
    q_{s}(t) & = f_{s}(t) + f_{\star}(t),\\
    \vdots \\
    q_{N_S}(t) & = f_{N_s}(t) + f_{\star}(t).
\end{aligned}
\label{eqn:full_model}
\end{equation}
Written in this way, it is easy to see that each spectrograph's time series contains an intrinsic drift GP alongside a common latent stellar variability GP shared across all instruments. Moreover, this framework allows us to extract each instrument's drift following the procedure outlined in \autoref{sec:gp_decomposition} without specifying or prescribing its functional form. The instrument model is implemented with independent kernels that behave as
\begin{equation}
    \begin{aligned}
    k_{\rm{s}}(t_i,t_j) &= k(t_i,t_j)\delta_{\rm{s}_i,\rm{s}_j}, \\
    k_{\rm{s}}(t_{i\ast},t_j) &= k(t_{i\ast},t_j), \\
    k_{\rm{s}}(t_i,t_{j\ast}) &= k(t_i,t_{j\ast}), \\
    k_{\rm{s}}(t_{i\ast},t_{j\ast}) &= k(t_{i\ast},t_{j\ast}),
    \end{aligned}
\end{equation}
where $k(t_i,t_j)$ is the kernel function of choice to describe the instrumental drifts (e.g., squared exponential), and $\delta_{\rm{s}_i,\rm{s}_j}$ is the Kronecker delta such that $\delta_{\rm{s}_i,\rm{s}_j} = 1$ if observation $i$ and $j$ are from the same spectrograph, and $\delta_{\rm{s}_i,\rm{s}_j}=0$ otherwise. This essentially acts as a mask in the covariance matrix of the observations, $\bm{K}_{ij}$, for each instrument GP; the mask does not act on the kernel functions if being computed for a test point ($\bm{K}_{\ast}$ or $\bm{K}_{\ast\ast}$). This kernel construction is relatively simple, but requires some care when computing the predictive means. In particular, when computing $\bm{K}_{\ast,s}^{\transpose}$ for a spectrograph component process, the test points $\{t_{\ast,1},t_{\ast,2},...t_{\ast,N}\}$ are then assumed to be test points for that specific spectrograph.

In summary, \autoref{eqn:full_model} is the full general GP model we use, with $N_l$ as the number of latent stellar components used in the model, and $N_s$ as the number of spectrographs in the data. We note that for the instrument components, each instrument GP $f_s(t)$ could itself be modeled as a sum of independent GPs akin to \autoref{eqn:stellar_gp} (e.g., to capture instrumental variations on multiple timescales. See discussion in \autoref{sec:discussion_inst_drift}.), but for this work, we will only ever use a single kernel process per instrument.

\section{Application to Simulated Datasets}\label{sec:synthetic_test}
\begin{figure}
    \centering
    \includegraphics[width=\linewidth]{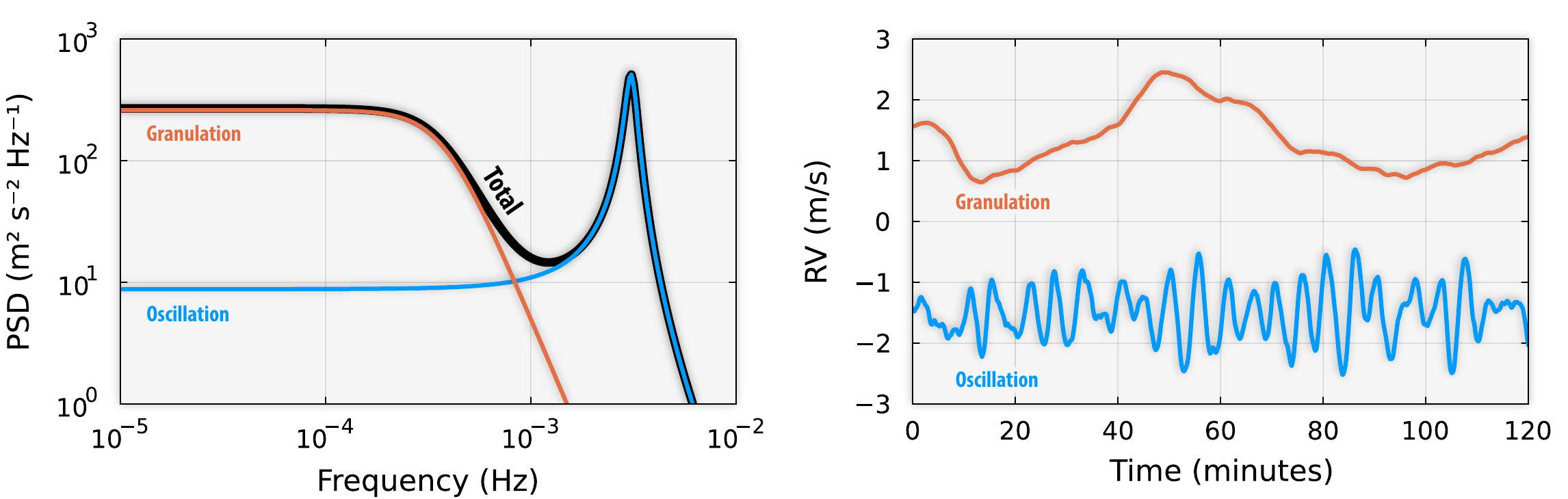}
    \caption{\emph{Left:} PSD for the solar variability kernels used in this work. We use a sum of a granulation kernel and an oscillation kernel. \emph{Right:} Sample draws for each of the stellar variability kernels (arbitrary offsets added for clarity.}
    \label{fig:PSD_sim}
\end{figure}

We now have ingredients to build a full GP model that can properly combine datasets from multiple instruments while accounting for the differing exposure times (both within and between instruments), and output the predictive mean of each stellar variability kernel as well as drifts between instruments. Before turning to real data, we wish to demonstrate these improvements on simulated datasets. Our simulated datasets will closely mimic real stellar data but differ in that we have knowledge of the true input signals. Our stellar variability kernels are SHO kernels with solar oscillation hyperparameters from \citet{Luhn2023} and solar granulation hyperparameters from \citet{O'Sullivan2025}:
\begin{equation}
\begin{aligned}
    \textbf{Oscillation:} \quad\quad &S_0 = 2.20 ~ \text{m}^2\text{s}^{-2} \,\text{Hz}^{-1} \quad\quad &\omega_0 = 0.0194~\text{rad}\,\text{s}^{-1}\quad\quad &Q =7.26 \\
    \textbf{Granulation:} \quad\quad &S_1 = 65.25 ~ \text{m}^2\text{s}^{-2} \,\text{Hz}^{-1} \quad\quad &\omega_1 = 0.0023~\text{rad}\,\text{s}^{-1}\quad\quad &Q \equiv 1/\sqrt{2} 
\end{aligned}
\label{eqn:solar_hypers}
\end{equation}
Note that the frequencies above are \emph{angular} frequencies $\omega = 2\pi \nu$ and the amplitudes $S$ are in normalized power units. Following \citet{Pereira2019}, we can convert the hyperparameters above to more intuitive $\sigma$ and $\lambda$ to describe the typical RMS amplitude of each kernel and the timescale over which each operates:
\begin{equation}
\begin{aligned}
    \textbf{Oscillation:} \quad\quad &\sigma_0 = 17 ~ \text{cm}\,\text{s}^{-1} \quad\quad &\lambda_0 = 5.39~\text{min}\\
    \textbf{Granulation:} \quad\quad &\sigma_1 = 33 ~ \text{cm}\,\text{s}^{-1}\quad\quad &\lambda_1 = 44.6~\text{min}\\
\end{aligned}
\label{eqn:solar_hypers_sigma_lambda}
\end{equation}
The PSDs for these kernels are shown in \autoref{fig:PSD_sim} along with a sample time series draw for each. These sample draws are evaluated at 1~s intervals and will be the ``truth'' astrophysical components for our simulation demonstration.

We now wish to sample these stellar signals as would be observed by different instruments. We choose 4 instruments (Insts. ``A'', ``B'', ``C'', and ``D'') with increasing exposure times chosen to closely match real Sun-as-a-star RV instruments ($t_{exp,A} = 12~\rm{s}$, $t_{exp,B} = 55~\rm{s}$, $t_{exp,C} = 178~\rm{s}$ and $t_{exp,D} = 300~\rm{s}$) so that the instruments A--D are ``KPF-like'' \citep{Rubenzahl2023}, ``NEID-like'' \citep{Lin2022}, ``EXPRES-like'' \citep{Llama2024}, and ``HARPSN-like'' \citep{Dumusque2015}, respectively (see also \citealt{Zhao2023} for a summary of typical values for NEID, EXPRES, and HARPS-N solar feeds). The readout times for each simulated instrument time series are also chosen to match their real-world counterpart ($t_{read,A} = 16~\rm{s}$, $t_{read,B} = 28~\rm{s}$, $t_{read,C} = 52~\rm{s}$ and $t_{read,D} = 26~\rm{s}$), where we have chosen the fast-readout mode of KPF \footnote{although fast-readout mode is not typical for KPF solar observations, it is useful to get a sense of ``best case'' sampling of stellar signals}. We compute the time series for each instrument by binning the true stellar signals over each observation's exposure. For our simulated case, we are most interested in demonstrating the performance of our kernel framework, so we limit the simulated time series to 2 hours. For this demonstration, we do not mimic the longitudinal spread of instruments and instead assume coverage from all 4 telescopes overlaps during this 3-hour spread. To summarize, our model for this demonstration contains $N_l = 2$ latent stellar processes and $N_s =4$ spectrographs for a total sum of 6 GP component processes.

\begin{figure}
    \centering
    \includegraphics[width=1\linewidth]{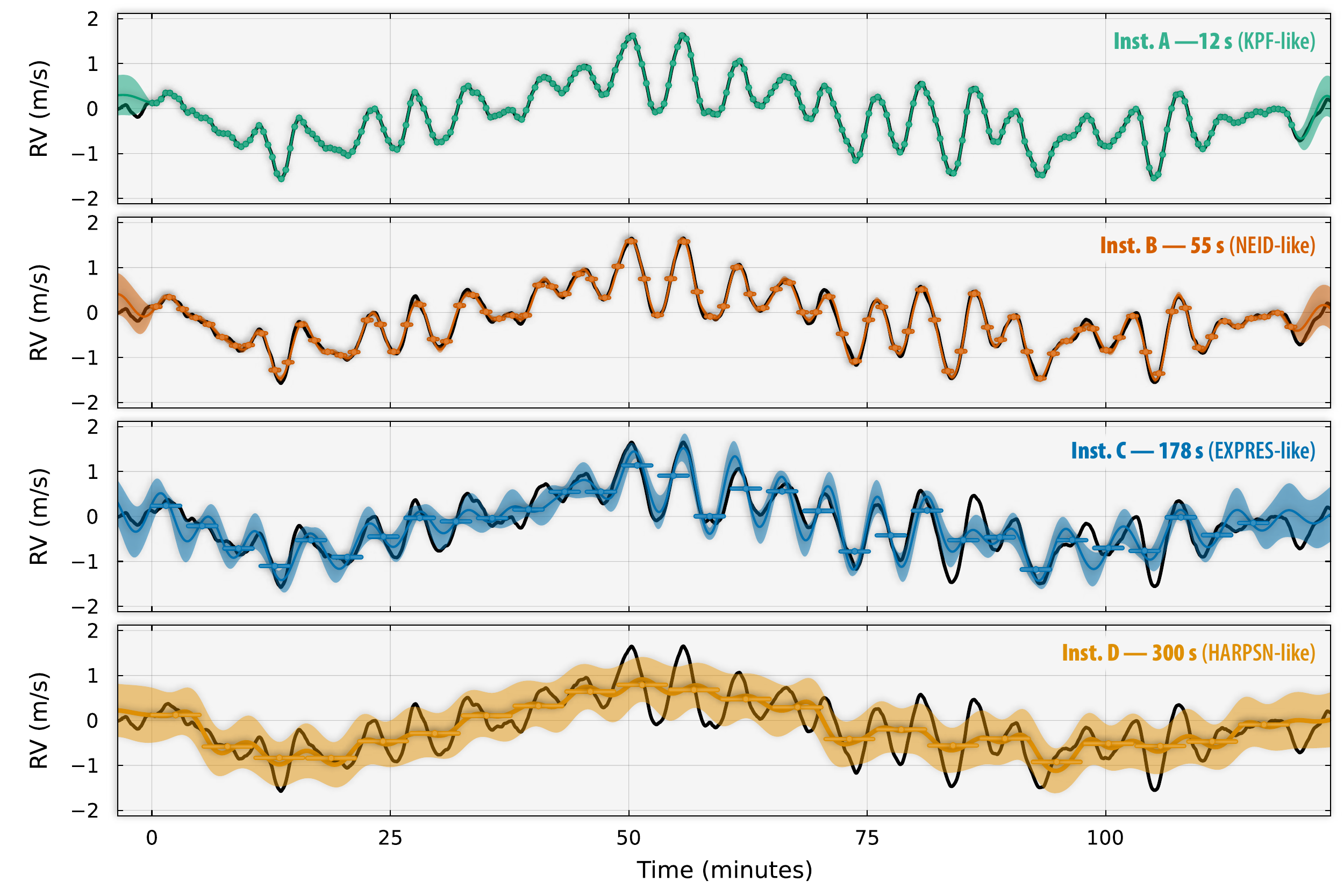}
    \caption{Simulated time series for our 4 representative instruments, assuming no instrumental noise. The colored horizontal lines indicate each observation's exposure time and lighter points display the observed RV centered within. The ``true'' stellar variability signal is shown in black in each panel (the sum of the components shown in \autoref{fig:PSD_sim}). For each instrument, we use the input GP model with fixed hyperparameters to compute the predictive mean for the latent signal (e.g., accounting for the exposure time of each observation), which is shown by the colored line in each panel, with the 1-$\sigma$ confidence region shaded. 
    }
    \label{fig:all_simulated_noise_free}
\end{figure}

\autoref{fig:all_simulated_noise_free} shows the noise-free observations from each of the 4 representative instrument cases and the resulting predictive mean after accounting for the exposure times with our GP model. While the KPF- and NEID-like instruments reliably trace the true signal, the EXPRES- and HARPSN-like instruments show clear effects from the extended exposure times that our GP model is able to account for. Typically, the predictive mean of noise-free data should pass directly through all observations, with uncertainties that shrink to zero at each observation---behavior that we expect to avoid with a model that is aware that each observation represents a time-averaged signal. Indeed, we see that our exposure-time accounting allows the predictive mean to be more extreme than the observations, especially evident when an observation is centered on an extrema, as seen in the dip before the 75-minute mark in the EXPRES-like time series. The HARPSN-like time series is a good demonstration of how the uncertainties do not collapse to zero at the time of observations and stay relatively wide throughout, reflecting the uncertainty from the family of functions whose binned average would result in the observed RV.

We note that the curves in \autoref{fig:all_simulated_noise_free} come from computing the predictive mean of the latent signal, which means evaluating the predictive mean at a finely sampled grid of points that are assumed to be instantaneous (i.e., exposure time = 0). We can alternatively compute the predictive mean of an \emph{integrated} signal, for any choice of exposure time. In our simulated datasets, each instrument observes with a fixed exposure time\footnote{This is true for the real life instruments that our representative sample is modeled after, with the exception of EXPRES, which exposes to a fixed SNR rather than a fixed exposure time}, and so we can separately compute the predictive mean for each instrument's exposure-averaged signal. For the case above of noiseless data, the predictive mean of an instrument-specific integrated (or \emph{exposure-averaged}) signal behaves as we would expect, i.e., it \emph{must} pass through each of that instrument's observations exactly with uncertainty that collapses to zero at its observations. Finally, the \emph{integrated} predictive mean can be computed for any combination of observation time and associated exposure time; this is particularly important for computing the predictive mean for each input observation, in order to produce meaningful residuals, as we describe below.

\begin{figure}
    \centering
    \includegraphics[width=\linewidth]{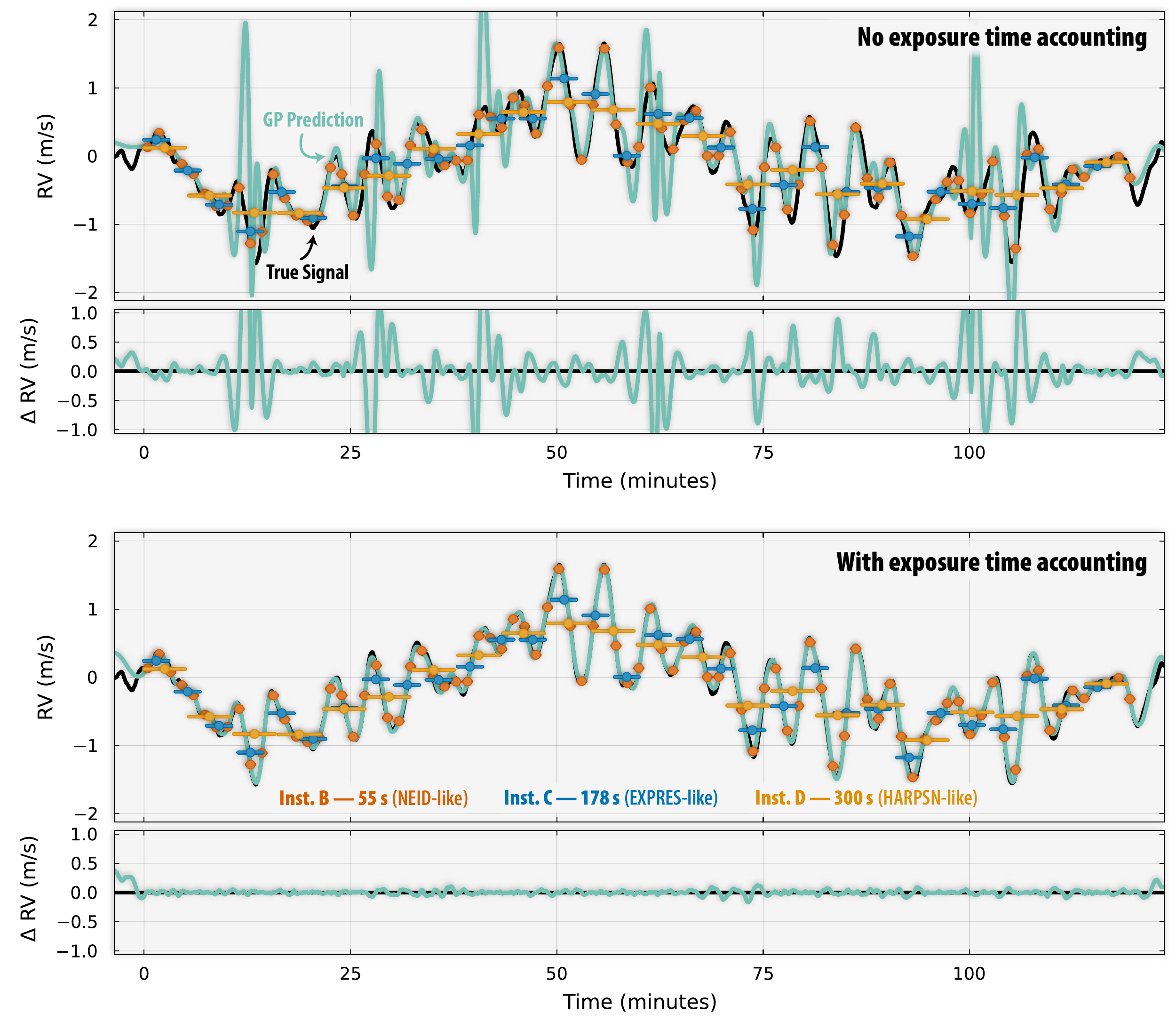}
    \caption{An example of computing the predictive mean on a combined dataset, combining the synthetic time series from Instruments B, C, and D. The top two panels show the predictive mean without exposure-time accounting (teal) as compared to the true signal (black), and the bottom two panels show the same predictive mean after properly accounting for the exposure times. For each case, the lower panel shows the residuals between the prediction and the true synthetic signal. Note the extreme departures when the model is forced to fit two nearby points if it does not account for the exposure times.}
    \label{fig:combined_noise_free}
\end{figure}

The exposure-time accounting is also important when combining data from multiple instruments. Without the accounting, the GP will assume all time series are true samples from the latent process, which leads to extreme departures when two observations with different exposure times occur very close to one another. We show an example of this effect in \autoref{fig:combined_noise_free} using the combined datasets of instruments B, C, and D. We again emphasize the distinction that the fits in \autoref{fig:all_simulated_noise_free} show the predicted \emph{latent} signal; the residuals between the points and the latent curve are not particularly meaningful, as it contains only the difference resulting from the exposure-averaging process. Instead, to compute meaningful residuals, one should compute the predicted mean using the input observation times and exposure durations; this process will result in the residuals between each observation and the appropriate exposure-averaged signal at that time. In the noiseless data case, each instrument's residuals are zero, demonstrating that the model accounts for the exposure times accurately.

We next add in photon noise to better simulate true observations. The photon noise is drawn from Normal(0,$\sigma_{RV}$) distributions with  $\sigma_{RV,A} = 0.21~\rm{m\,s^{-1}}$ \footnote{The precision improvement from 0.3~\ms{} in \citet{Rubenzahl2023} is a result of changes to how KPF SoCal data is now taken: the exposure times have increased from 5 to 12~s, and fewer orders are masked out. 0.21~\ms{} represents the current typical single-measurement precision for 12~s KPF SoCal observations.}, and from \citet{Zhao2023} for $\sigma_{RV,B} = 0.16~\rm{m\,s^{-1}}$, $\sigma_{RV,C} = 0.19~\rm{m\,s^{-1}}$, and $\sigma_{RV,D} = 0.26~\rm{m\,s^{-1}}$. In general, we find that the differences between accounting for and not accounting for the exposure times are less extreme when noisy observations are used. This is because the model that does not account for exposure times now has some additional flexibility to account for the ``discrepant'' exposure-averaged observations, which it can attribute to the observational uncertainty. The observations with the longest exposure times will always observe a latent signal that has lower RMS amplitude as a result of the exposure averaging. Thus, the model that does not account for exposure times will split the difference between the instruments with different exposure times and lead to a predictive mean that underpredicts the true amplitude. However, the instruments with longer exposure times will typically have lower cadence, and therefore be underweighted relative to the higher-cadence dataset. Therefore, when not accounting for the exposure times, the fits typically end up latching on to the signal from the higher-cadence datasets, reducing the amplitude of the predicted signal slightly to reconcile the exposure-averaging. The degree of amplitude reduction will depend upon the relative number of observations between instruments with longer and shorter exposures, the degree to which each time series overlaps, and the exact uncertainties for each instrument.

\begin{figure}
    \includegraphics[width=\textwidth]{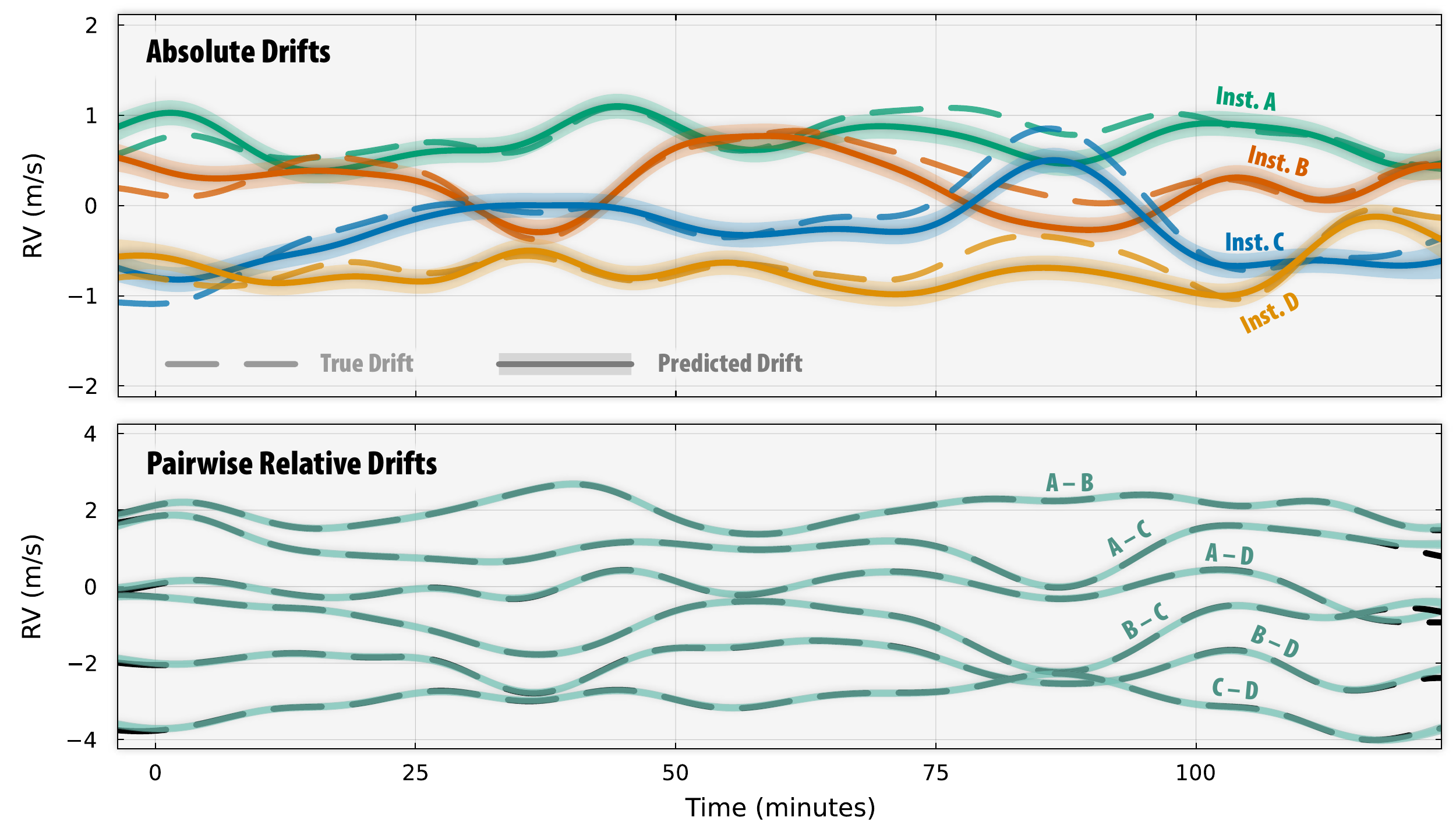}
    \caption{Recovery of instrument drifts from combining noise-free observations from 4 instruments with a model that includes both stellar signals and instrument drifts. Top panel: Recovered, or ``predicted'' instrument drifts (solid lines with 1-$\sigma$ confidence regions) compared to the true synthetic drifts for each instrument in the drift model case. Recovered drifts are overall well correlated with the true drifts, however, large departures exist, particularly evident between 75 and 100 minutes in this example. Bottom panel: Instead, pairwise relative drifts (e.g., differential drifts between instrument pairs in the top panel) are extremely well constrained, with the recovered pairwise relative drifts (semi-transparent light teal line) only differing from the true relative drifts in the edges of the time series beyond the observations ($t < 0$ min. or $t > 120$ min.); the residuals between the true relative drift and the recovered relative drift for $0 < t< 120$ min has RMS $< 1.5$ \cms\ for all curves.}
    \label{fig:drift_comparison}
\end{figure}

Next, we test our recovery of instrument drifts with the drift model described in \autoref{sec:instrument_drift}. We inject randomly-drawn synthetic drifts for each instrument, each drawn from a squared exponential (SE) kernel with lengthscale 7.2 minutes and amplitude 30~\cms. As a simple test, we use the input hyperparameters as the hyperparameters used to condition the model; in practice, one would not know the underlying hyperparameters a priori and would need recover these via a fit, e.g., maximum likelihood estimation. Our simple test is meant to demonstrate our ability to recover instrument drifts under idealized conditions. We find that even in these conditions, absolute drifts are not well recovered; the top panel of \autoref{fig:drift_comparison} shows that while the recovered drifts are decently correlated with the true drifts, there is considerable tradeoff between the different instruments. We conclude that even with perfect overlapping coverage between 4 instruments, there will always be ambiguity about which instrument is drifting in which direction at any given time. However, as we show in the bottom panel of \autoref{fig:drift_comparison}, the \emph{relative} drifts between any two instruments is constrained extremely well. Thus, we conclude that our instrument drift model is successful at distinguishing combined instrumental drifts from stellar signals; we refer to these drifts as ``pairwise relative drifts.'' Unfortunately, this suggests that the recovered absolute drifts of any simultaneous instruments will be strongly correlated (often anticorrelated as the model prefers to split the difference with each instrument drifting in opposite directions), significantly complicating any efforts to tie the recovered drift model to missing physics or unused telemetry in that instrument's existing drift corrections as part of its data reduction pipeline. However, in a scenario where the true drifts have significantly different hyperparameters between each instrument (e.g., different drift timescales), the GP model will more easily identify each instrument's absolute drift. In the limit of many instruments, the model will likewise gain confidence in assigning drift signals to the appropriate offending instrument(s).

\section{Application to Solar Datasets}\label{sec:real_test}
We now wish to apply these kernels to real solar data. We will examine two test cases: (1) a sample day of KPF \& NEID solar data with significant overlap, and (2) one month of solar data from HARPS, HARPS-N, EXPRES, and NEID as presented in \citet{Zhao2023}.

\subsection{KPF \& NEID Solar Data}
\begin{figure}
    \centering
    \includegraphics[width=\linewidth]{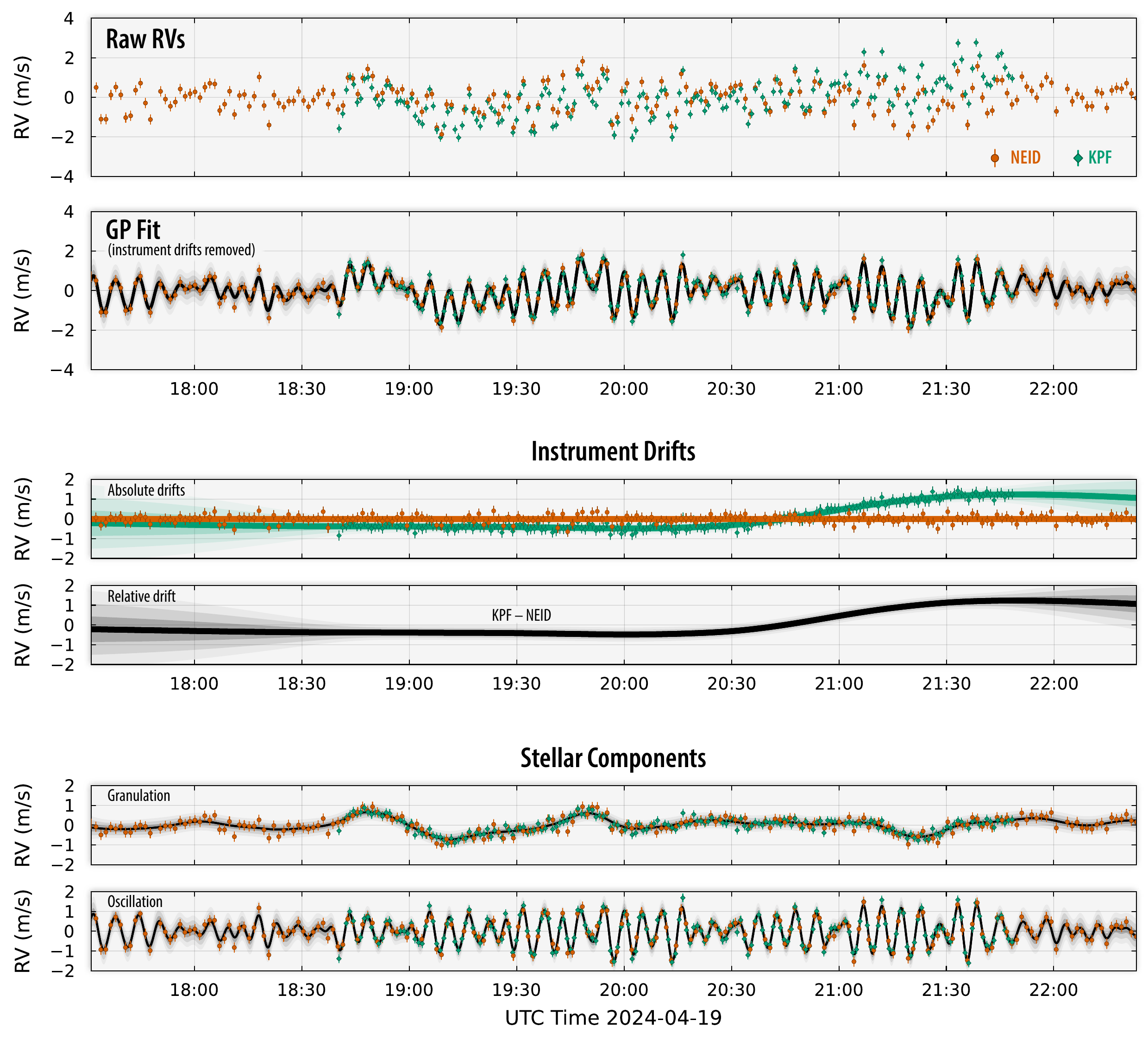}
    \caption{Combined GP model fitting between overlapping NEID and KPF observations taken on 2024, April 19. The top panel shows the raw RVs from KPF (green diamonds) and NEID (orange circles) observations after mean subtraction. The second panel shows the result of our GP model, where the black line shows the stellar components (sum of granulation and oscillations), and the data have been corrected using the predicted GP drift component for each instrument.  The next two panels show the absolute NEID and KPF drift components and their differential (``relative'') drift, respectively; the relative drift is never more than 50~\cms\ in either direction between the two instruments. The final two panels show the individual stellar components, granulation and oscillations, respectively. Note that the amplitudes of the variability are significantly larger than the recovered instrumental drifts.}
    \label{fig:neid_kpf_april_2024}
\end{figure}

As a first case, we choose publicly available NEID and KPF data, selecting 2024 April 19 as a date with substantial overlap between the two instruments; morning observations from KPF overlap with afternoon observations from NEID. Both locations saw fully clear skies during their respective observations as monitored by each instrument's pyrheliometer and quantified by the clearness metric of \citet{Rubenzahl2023}. We first subtract the mean from each data set to ensure that each is a zero-mean time series. We construct a 4-component GP model containing $N_s= 2$ spectrographs (using squared exponential kernels for both KPF and NEID), and $N_l=2$ latent stellar components (a granulation and an oscillation kernel). We fix the hyperparameters of the stellar components to the solar values used for the synthetic tests above (\autoref{eqn:solar_hypers} and shown in \autoref{fig:PSD_sim}), and perform a fit to the instrument drift kernels using the nested sampling routine found in \texttt{NestedSamplers.jl} \citep{nestedsamplers}. We use uniform priors for the amplitude, $U(0.1,10)$~\ms, and lengthscale, $U$(8~s,1~day) and sample the prior space using the random walk algorithm with multi-ellipsoidal bounds \footnote{The multi-ellispoidal bounds improve the sampling efficiency in the case of multi-modal posteriors by bounding each cluster in (potentially overlapping) ellipsoids. See \citet{Mukherjee2006} for a basic description of the algorithm; the algorithm in \texttt{NestedSamplers.jl} follows that of \texttt{MultiNest} as described in \citet{Feroz2008} and with an updated multi-ellipsoidal sampling algorithm in \citet{feroz2009}.} and 200 live points; we impose a cutoff threshold of $\Delta \log{Z}= 0.1$. The resulting posterior gives a KPF instrument drift kernel with amplitude 95~\cms\ and timescale 1.43~hr; the posterior for the NEID drift kernel has amplitude 15~\cms\ and timescale of 2.6 hours. Using these values, we compute the predictive mean of the combined KPF and NEID data set and each component kernel; \autoref{fig:neid_kpf_april_2024} shows the predictive means for stellar and instrumental components from this joint fit to the NEID and KPF time series. In this case, the raw NEID time series is rather well behaved, and the resulting NEID drift component is essentially flat. The KPF drift component smoothly identifies what we can see by eye in the raw RVs: relative to NEID, the KPF RVs drift slightly over the course of the overlapping sequence from slightly below NEID to about 1~\ms\ above NEID at the end of the KPF sequence. The uptick attributed to KPF drift near UT 20:30 corresponds to the time of the daily liquid nitrogen fills for KPF, which on this day began just after UT 20:30. To test that the recovered instrument drifts are not strongly dependent on the kernel function used, we constructed a second model where the instrument drifts are modeled as a granulation-like kernel, chosen to emulate a kernel with constant power below a certain cutoff frequency. We fit the hyperparameters in the same way, and again find that the NEID drift model is lower amplitude and operates on longer timescales than the KPF instrument drift model. The predicted drifts are indistinguishable from the case where we modeled the instrument drifts using a squared exponential kernel. We also tested a case where we randomly injected a spurious RV ``bump'' into the NEID RVs. After refitting the instrument drift hyperparameters, we successfully recover the injected bump as a NEID drift; in this case, some additional variability originally attributed to KPF ends up being attributed to NEID. However, the relative drift is essentially identical except for the added bump.

\subsection{Solar Data from The Extreme Stellar Signals Project}
With our second example, we wish to demonstrate how this framework can be used for combined datasets over long timescales. We turn to the solar data from the Extreme Stellar Signals Project \citep{Zhao2023}, which combined solar RVs from HARPS, HARPS-N, EXPRES, and NEID over a 28-day span from 2021 May 26 to 2021 June 21. The dataset contains 4692 HARPS observations with 30 s exposures, 1397 HARPS-N observations with 300 s exposures, 1459 EXPRES observations with median exposure time of 177.5 s (EXPRES exposes to a fixed SNR rather than fixed time), and 3617 NEID observations with 55 s exposures; in total, there are 11,165 observations. Given the $N^3$ scaling for GP evaluations, this dataset is too large at present to perform hyperparameter fitting as in the nested sampling routine in the previous section. With our existing unoptimized methods and setup, just conditioning the model on the data (representing the equivalent time for a single iteration) takes more than 20 minutes. However, from our exploration in the KPF \& NEID example, we determined that the instrument drifts are consistent between different kernels, indicating there is some flexibility in our choice of kernel. With this in mind, we choose to use a fixed model and simply condition it on the data, saving more robust model fitting for future implementations of our framework with improved scaling performance.

Because the combined dataset now spans several weeks, we need to include additional stellar components to our model. In addition to the  granulation and oscillation kernels we have used throughout, we include an additional granulation-like ($Q\equiv 1/\sqrt{2}$) SHO kernel with timescale 1.05 days and amplitude 0.72~\ms\ based on the ``supergranulation'' component in \citet{O'Sullivan2025}, and an additional SHO kernel with timescale 12 days, amplitude 2.5 \ms\ and oscillation strength Q = 6. This last kernel represents a moderately coherent oscillator with period near half of the solar rotation period, to mimic the impact of rotationally modulated spots and faculae. For our dataset that spans 30 days, such a simplistic model is sufficient to capture the rotational signal. Each instrument is assigned a drift kernel of a squared exponential (SE) with 25 \cms\ amplitude and 36-minute timescale\footnote{We note again that the predicted instrument drifts are insensitive to the form of the kernel, as we obtain similar results with a granulation-like component. The timescales and amplitudes were chosen after several iterations. The timescale is short enough to capture observed departures from each other while still long enough to avoid significant overfitting; we have chosen the amplitude in a similar manner. The values chosen here are purely representative---with a framework that can solve the GP scaling issues, all amplitudes and timescales should be treated as free parameters.}. In summary, following the same $\sigma$ and $\lambda$ notation of \autoref{eqn:solar_hypers_sigma_lambda}, our GP model has $N_l = 4$ latent stellar processes and $N_s=4$ spectrograph processes:
\begin{equation}
\begin{aligned}
    \textbf{Oscillation (SHO):} \quad\quad &\sigma_0 = 0.17 ~ \text{m}\,\text{s}^{-1} \quad\quad &\lambda_0 &= 5.39~\text{min} \quad\quad &Q = 7.26 \\
\    
    \textbf{Granulation (SHO):} \quad\quad &\sigma_1 = 0.33 ~ \text{m}\,\text{s}^{-1}\quad\quad &\lambda_1 &= 44.6~\text{min} \quad\quad &Q \equiv 1/\sqrt{2}\\
\    
    \textbf{Supergranulation (SHO):} \quad\quad &\sigma_2 = 0.72 ~ \text{m}\,\text{s}^{-1}\quad\quad &\lambda_2 &= 25.3~\text{hr} \quad\quad &Q \equiv 1/\sqrt{2}\\
\    
    \textbf{Rotation (SHO):} \quad\quad &\sigma_3 = 2.5 ~ \text{m}\,\text{s}^{-1}\quad\quad &\lambda_3 &= 12~\text{day} \quad\quad & Q= 6\\
\    
    \textbf{Inst}_{\textbf{HARPS}}\textbf{ (SE):} \quad\quad &\sigma_{4} = 0.25 ~ \text{m}\,\text{s}^{-1}\quad\quad &\lambda_4 &= 36~\text{min} \quad\quad &\\
\    
    \textbf{Inst}_{\textbf{HARPS-N}}\textbf{ (SE):} \quad\quad &\sigma_{5} = 0.25 ~ \text{m}\,\text{s}^{-1}\quad\quad &\lambda_5 &= 36~\text{min} \quad\quad &\\
\    
    \textbf{Inst}_{\textbf{EXPRES}}\textbf{ (SE):} \quad\quad &\sigma_{6} = 0.25 ~ \text{m}\,\text{s}^{-1}\quad\quad &\lambda_6 &= 36~\text{min} \quad\quad &\\
\    
    \textbf{Inst}_{\textbf{NEID}}\textbf{ (SE):} \quad\quad &\sigma_{7} = 0.25 ~ \text{m}\,\text{s}^{-1}\quad\quad &\lambda_7 &= 36~\text{min} \quad\quad &\\
\end{aligned}
\end{equation}

\begin{figure}
    \includegraphics[width=\textwidth]{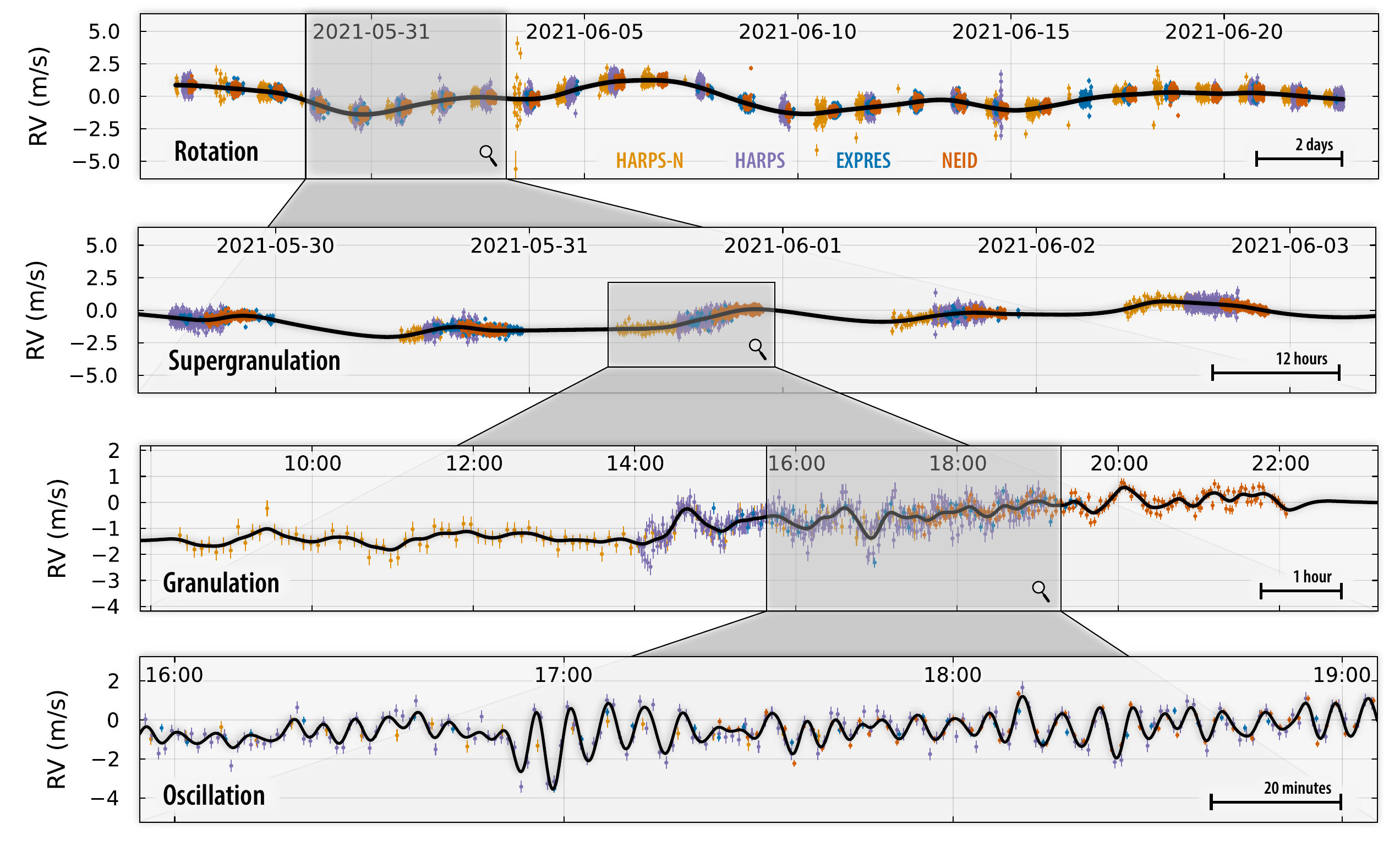}
    \caption{GP predictive means for each stellar component after combining HARPS-N (yellow), HARPS (purple), EXPRES (blue), and NEID (orange) solar data from the ESSP dataset \citep{Zhao2023}; successive panels zoom in from the full dataset (28 days) to a single portion of one day (3 hours). We have subtracted the instrument drift component from each dataset. In each panel, the GP prediction in black shows the sum of the listed component and all components with longer timescales, whereas the data points plot the drift-corrected RVs after subtracting the contribution from components with shorter timescales. For example, in the ``Supergranulation'' panel, the black line shows the sum of supergranulation and rotation, while the plotted data points have had the granulation and oscillation components subtracted out.}
    \label{fig:ESSP_combo}
\end{figure}

We show a summary of the stellar signals present in the combined dataset after using our GP framework in \autoref{fig:ESSP_combo}. With the conditioned model, we can isolate any combination of stellar and/or instrumental components for further study, though we leave such deeper dives for a further study. For now, we demonstrate that our model can handle the intricacies of combining multiple datasets and leveraging the additional constraints they provide. In the future, we expect an improved framework will allow us to perform hyperparameter tuning and proper model comparison to fully model the solar variability.

\section{Broader Applications of Exposure-integrated GP Kernels }\label{sec:broader_applications}

The GP integration framework we presented in \autoref{sec:integration_framework} was motivated by time domain observations that have been binned over a given exposure time. However, this approach is not limited to time-domain applications; it applies to any binned observation where the bin size is an appreciable fraction of the lengthscale of an underlying latent process. A recent example in galactic astronomy used integrated GP kernels to map interstellar dust, where the integrated kernels accounted for the unobserved line-of-sight variations in dust density.

Another common astronomical measurement is an observation made over a discrete bandpass, where the observed flux represents the integrated flux across the band's wavelength range. For example, transit spectroscopy requires a planet's transit to be observed in multiple bands simultaneously to measure the varying optical depth of the annulus of planet's atmosphere as a function of wavelength to detect absorption features and the presence of clouds or hazes \citep{Charbonneau2002,Sing2011}. Analogous to our exposure time examples, each measurement of the planet's transmission spectrum is a measurement of the binned version of the true latent signal. In atmospheric retrieval, the latent signal is often fit using a parametric forward model with free parameters that describe the atmospheric profile and the radiative transfer. With our integrated GP framework, it would be straightforward to build in additional spectral variability as a Gaussian process signal that has been binned over the bandpass. Such an approach may offer additional improvements over current approaches to combining datasets from instruments with overlapping bandpasses, e.g., combining spectra from the four overlapping JWST observing modes as in \citet{Carter2024}.

\section{Discussion}\label{sec:discussion}
\subsection{Applications to Planet-hunting Datasets}
The benefits of the integrated kernel framework come primarily from combining datasets with significant overlapping measurements; furthermore, the effects of differing exposure times are negligible for long-timescale sources such as rotational modulation. We do not expect this framework to be immediately applicable to existing planet-hunting datasets for those reasons, as survey-cadence observations are often spaced such that consecutive visits are essentially independent for the kernels where thex integrated form is relevant (oscillations and granulation); the contribution of these sources to a GP model can be sufficiently captured with the standard ``jitter'' term alongside a rotation-based quasi-periodic kernel. A robust study of alternative survey strategies (e.g., high cadence nightly observations to model and remove oscillations and granulation) is beyond the scope of this work, but we suggest that any such studies should make use of the integrated framework presented here. This framework is instead most beneficial in the case of isolating distinct sources of stellar variability in high-cadence datasets, such as the solar data streams used in our examples, or in unique cases of coordinated efforts uniquely designed to study stellar variability. 

\subsection{Scaling and Non-stationarity of Integrated Kernels}
Our integrated GP framework is currently limited by the GP scaling issue (i.e., computations and memory scale cubically and quadratically with the number of data points, respectively) when applied to extremely large datasets ($N>3000$), which prevents us from performing MCMC fits to tune the GP kernel hyperparameters. Some workarounds exist, but most are inexact, only inferring the relevant covariance matrix structure, e.g., inducing points \citep{Wilson2015,Banerjee2008,Finley2009,Miller2022}; further, such methods are not ideal in the case of stellar variability, as the relevant timescales span many orders of magnitude and cannot handle exposure overlap. The methods that retain the exact structure of the covariance matrix require specific family of kernels \citep[e.g., ``celerite'']{ForemanMackey2017} or leverage semi-separable and ``LEAF'' structure \citep{Delisle2020,Delisle2022}; while the latent kernels we use have a \texttt{celerite} form, the integrated form of the kernels are no longer stationary\footnote{We note that in the limiting case of non-overlapping \emph{and constant exposure times}, the integrated kernel is stationary and \texttt{celerite} form can be used}, and instead are a function of $|t-t'|$ and the two exposure times $\delta_1,\delta_2$. Further, overlapping observations from multiple instruments break the ``LEAF'' structure of the matrix, as gaps can enter into the covariance matrix depending on whether subsequent observations have overlap or not. 

However, the integrated kernels are only semi-non-stationary; that is, while the covariance does not depend \emph{only} on $|t-t'|$, it is still invariant to $t \leftrightarrow t'$ (i.e., the kernels are not time dependent). It is possible that the resulting covariance matrix is semi-separable and there exist some creative solutions that allow us to use \texttt{celerite} kernels. Another promising solution could make use of sparsity-discovering kernels \citep{Noack2023}, since the ``complicated'' part of the covariance matrix (exposure overlaps) are very close to the diagonal. We leave these these possible GP approaches to future studies.

Fortunately, an alternative but equivalent representation of GPs does afford a scalable implementation of exposure-integrated measurements (including overlap). \citet{Rubenzahl2026} developed the equivalent state-space model to our problem statement, which is solved by an augmented state and matching Kalman filter and RTS smoother; the solution has $\mathcal{O}(N)$ complexity and is parallelizable. With a scalable framework in place, we are extending these investigations in upcoming studies of stellar variability and instrument drifts in a more robust way and on longer timescales \citep{Rubenzahl2026}.

\subsection{More Complicated Drift Frameworks}\label{sec:discussion_inst_drift}
For our demonstrations, the drift framework contained simple kernels specified by a single timescale. In reality, the instrument drifts will vary on many timescales, and with more complicated behavior. For instance, daytime solar observations often occur soon after daily liquid nitrogen fills as part of regular instrument operations and calibration sequences. As a result, the daily RVs often contain discontinuities or ramps near the same time of each day. To account for this and other effects, the instrument kernels will likely need to be a sum of multiple kernels that operate on different timescales and/or contain change points to handle discontinuities/ramps. We would also like to investigate the structure of differential extinction in various solar data sets, which should similarly be a shared latent (chromatic) signal across all instruments, with a time lag to account for each observing site's geographic location. Such a model might instead use a kernel that maps each timestamps to disc-averaged airmass. This is an easy kernel to implement in the general GP framework, but may not be possible with a state-space model which requires sortable (i.e. monotonic) coordinates (e.g., a timeseries).

\section{Summary and Conclusions}\label{sec:conclusions}
In this paper we have presented a GP framework that (1) accounts for the integrated effects of discrete exposure times and overlapping observations, (2) decomposes the predictive mean of a multi-component model into its constituent kernels, and (3) incorporates drifts between multiple instruments for combined datasets. The framework was motivated by the growing number of EPRV instruments with solar feeds, each with differing and overlapping exposure times and relative drifts; however, it could be applied to any dataset where the observations have been binned over a lengthscale that is comparable to the relevant GP lengthscales. 

We constructed the integrated GP kernels using three ingredients. The first was the traditional instantaneous kernel that describes the covariance between any two instantaneous points (the latent function). The second was the singly integrated form of the kernel to describe the covariance between an instantaneous points and a binned observation. The final piece was the doubly integrated kernel for the covariance between any two binned observations. We further demonstrated how these integrals behave for stationary kernels, where the sign of $|t-t'|$ flips across the integral: we define each of the necessary integrals in the case of perfectly overlapping observations and separate observations and show that any two partially overlapping observations can be decomposed into sums of integrals of those forms. Together, these pieces allow one to properly account for the covariance between binned observations and evaluate the predictive mean of the latent function. 

We then derived the analytic expressions for the integrated kernels for the case of the simple harmonic oscillator, a kernel that is commonly used for modeling short-timescale solar variability, e.g., granulation and oscillations. 

Next, we reviewed the formalism for evaluating the predictive mean of each component of a multi-component GP model. We then introduced a simple method for accounting for drifts between overlapping datasets from multiple instruments by using independent kernels such that each kernel function only computes covariances between observations from the same instrument. 

With the complete framework, we first tested it on simulated noise-free data sets generated to mimic the cadence and exposure of real EPRV solar feeds from KPF, NEID, EXPRES, and HARPS-N. We show that not accounting for the exposure times leads to erroneous predicted signals; the errors are largest when two observations occur very close together with different exposure times. When including instrument drifts, we find that absolute drifts are recovered reasonably well, but with periods of ambiguity where the drift is incorrectly attributed to a different instrument. However, the relative drift between any instrument pair (which we refer to as the ``pairwise relative drift'') is recovered to within 1.5~\cms\ in the noise-free scenario. 

Next, we applied our framework to two real datasets. We started with a single day of overlapping RVs from KPF and NEID. We then applied it to a month of data from HARPS-N, HARPS, EXPRES, and NEID data from the Extreme Stellar Signals Project \citep{Zhao2023}. In both cases, we were able to compute predictive means of individual stellar components and verify that the framework properly handled the overlapping exposures and instrumental drifts. With this modeling framework in hand, and scalable approaches forthcoming \citep{Rubenzahl2026}, future studies will explore the characteristics of stellar variability and instrument drift at finer resolution and longer timescales.

\begin{acknowledgements}
We thank the anonymous referee for thoughtful and constructive feedback that substantially improved the clarity and readability of this manuscript. We are also grateful to Eric Ford for insightful advice on notation.

J.K.L. is supported by an appointment to the NASA Postdoctoral Program at the NASA Jet Propulsion Laboratory, administered by Oak Ridge Associated Universities under contract with NASA.

The research was carried out, in part, at the Jet Propulsion Laboratory, California Institute of Technology, under a contract with the National Aeronautics and Space Administration (80NM0018D0004) and funded through the President’s and Director’s  Research \& Development Fund Program.

Support for this work was provided by NASA through the NASA Hubble Fellowship grant HST-HF2-51569 awarded by the Space Telescope Science Institute, which is operated by the Association of Universities for Research in Astronomy, Inc., for NASA, under contract NAS5-26555.

The HELIOS solar telescope is located at the ESO 3.6m Telescope facility at the La Silla Observatory in Chile, and is connected to the HARPS and NIRPS spectrographs.  HELIOS (PI: Xavier Dumusque \& Pedro Figueira) was funded through Portuguese FCT funding and the Branco Weiss fellowship. 

The HARPS-N solar telescope is located at the TNG facility at the Roque de los Muchachos Observatory in La Palma, Spain, and is connected to the HARPS-N spectrograph.  The HARPS-N solar telescope (PI: Xavier Dumusque \& David Phillips) was funded with support from Smithsonian R\&D internal funding and the Branco Weiss fellowship.

The HARPS-N project was funded by the Prodex Program of the Swiss Space Office (SSO), the Harvard University Origin of Life Initiative (HUOLI), the Scottish Universities Physics Alliance (SUPA), the University of Geneva, the Smithsonian Astrophysical Observatory (SAO), the Italian National Astrophysical Institute (INAF), University of St. Andrews, Queen’s University Belfast, and University of Edinburgh.  We thank the HARPS-N solar team and TNG staff for processing the solar data and maintaining the solar telescope.

Maintenance and upgrade of both the HELIOS and the HARPS-N telescope was possible thanks to support from the European Research Council (ERC) under the European Union’s Horizon 2020 research and innovation programme (grant agreement SCORE No 851555) and from the Swiss National Science Foundation under the grant SPECTRE (No 200021\_215200). 

These results made use of the Lowell Discovery Telescope at Lowell Observatory. Lowell is a private, non-profit institution dedicated to astrophysical research and public appreciation of astronomy and operates the LDT in partnership with Boston University, the University of Maryland, the University of Toledo, Northern Arizona University and Yale University.

The EXPRES team acknowledges support for the design and construction of EXPRES from NSF MRI-1429365, NSF ATI-1509436 and Yale University. EXPRES PI Debra Fischer gratefully acknowledges support to carry out this research from NSF 2009528, NSF 1616086, NSF AST-2009528, the Heising-Simons Foundation, and an anonymous donor in the Yale alumni community.

Some of the data presented herein were obtained at
the W. M. Keck Observatory, which is operated as a
scientific partnership among the California Institute of
Technology, the University of California, and the National Aeronautics and Space Administration. The Observatory was made possible by the generous financial
support of the W. M. Keck Foundation. This research
has made use of the Keck Observatory Archive (KOA),
which is operated by the W. M. Keck Observatory and
the NASA Exoplanet Science Institute (NExScI), under
contract with the National Aeronautics and Space Administration. The authors wish to recognize and acknowledge the very significant cultural role and reverence that the summit of Maunakea has always had within the Native Hawaiian community. We are most fortunate to have the opportunity to conduct observations from this mountain.

This paper contains data taken with the NEID instrument, which was funded by the NASA-NSF Exoplanet Observational Research (NN-EXPLORE) partnership and built by Pennsylvania State University.  

NEID is installed on the WIYN telescope, which is operated by the NSF's National Optical-Infrared Astronomy Research Laboratory (NOIRLab).  The NEID archive is operated by the NASA Exoplanet Science Institute at the California Institute of Technology.  Part of this work was performed for the Jet Propulsion Laboratory, California Institute of Technology, sponsored by the United States Government under the Prime Contract 80NM0018D0004 between Caltech and NASA.

This paper is based in part on observations at Kitt Peak National Observatory, NSF’s NOIRLab, managed by the Association of Universities for Research in Astronomy (AURA) under a cooperative agreement with the National Science Foundation. The authors are honored to be permitted to conduct astronomical research on Iolkam Du’ag (Kitt Peak), a mountain with particular significance to the Tohono O’odham.

Deepest gratitude to Zade Arnold, Joe Davis, Michelle Edwards, John Ehret, Tina Juan, Brian Pisarek, Aaron Rowe, Fred Wortman, the Eastern Area Incident Management Team, and all of the firefighters and air support crew who fought the recent Contreras fire. Against great odds, you saved Kitt Peak National Observatory.

The Center for Exoplanets and Habitable Worlds and the Penn State Extraterrestrial Intelligence Center are supported by Penn State and the Eberly College of Science.  Computations for this research were performed on the Penn State’s Institute for Computational and Data Sciences' Advanced CyberInfrastructure (ICDS-ACI).  This content is solely the responsibility of the authors and does not necessarily represent the views of the Institute for Computational and Data Sciences.

\end{acknowledgements}


\bibliography{refs}{}
\bibliographystyle{aasjournalv7}



\end{document}